\documentclass[prd,12pt,tightenlines, aps, letterpaper,amsmath,amssymb,preprintnumbers,showpacs,superscriptaddress,floatfix,longbibliography,nofootinbib]{revtex4-1}
\usepackage{xspace}
\usepackage{graphicx,comment}
\usepackage{tikz}
\usepackage{subfigure}
\allowdisplaybreaks % let equations split over multiple pages
\usepackage[hypertexnames=false]{hyperref}
\hypersetup{
    colorlinks=true,       % false: boxed links; true: colored links
    linkcolor=blue,          % color of internal links
    citecolor=blue,        % color of links to bibliography
    filecolor=blue,      % color of file links
    urlcolor=blue           % color of external links
}
\usepackage{slashed}
\usepackage{mathtools}

\newcommand{\hethree}{{}^3\text{He}}
\newcommand{\hthree}{{}^3\text{H}}
\newcommand{\ket}[1]{\left| #1 \right>} % for Dirac bras
\newcommand{\mbraket}[3]{\left< #1 \vphantom{#2#3} \right|
 #2 \left| #3 \vphantom{#1#2} \right>} % for Dirac matrix elements

\begin{document}

\begin{figure}
  \vskip -1.5cm
  \leftline{\includegraphics[width=0.15\textwidth]{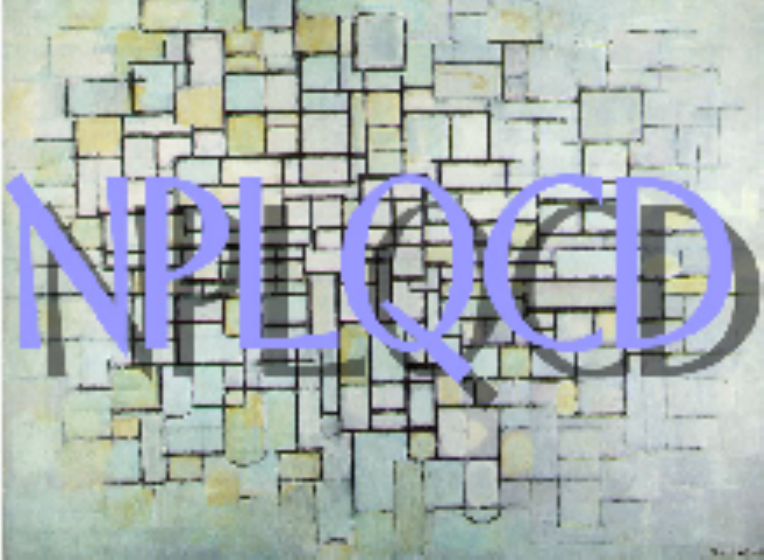}}
\end{figure}

\title{The axial charge of the triton from lattice QCD}

\author{Assumpta~Parre\~no}
\affiliation{Departament de F\'{\i}sica Qu\`{a}ntica i Astrof\'{\i}sica and Institut de Ci\`{e}ncies del Cosmos,	Universitat de Barcelona, Mart\'{\i} Franqu\`es 1, E08028, Spain}

\author{Phiala E. Shanahan } \affiliation{
	Center for Theoretical Physics, 
	Massachusetts Institute of Technology, 
	Cambridge, MA 02139, USA}
\affiliation{The NSF AI Institute for Artificial Intelligence and Fundamental Interactions}

\author{Michael~L.~Wagman}
\affiliation{
	Center for Theoretical Physics, 
	Massachusetts Institute of Technology, 
	Cambridge, MA 02139, USA}
\affiliation{Fermi National Accelerator Laboratory, Batavia, IL 60510, USA}

\author{Frank Winter}
\affiliation{Jefferson Laboratory, 12000 Jefferson Avenue, 
	Newport News, VA 23606, USA}
	
\author{Emmanuel Chang} \noaffiliation

\author{William Detmold} 
\affiliation{
	Center for Theoretical Physics, 
	Massachusetts Institute of Technology, 
	Cambridge, MA 02139, USA}
\affiliation{The NSF AI Institute for Artificial Intelligence and Fundamental Interactions}

\author{Marc Illa}
\affiliation{Departament de F\'{\i}sica Qu\`{a}ntica i Astrof\'{\i}sica and Institut de Ci\`{e}ncies del Cosmos,	Universitat de Barcelona, Mart\'{\i} Franqu\`es 1, E08028, Spain}

\collaboration{NPLQCD collaboration}

\begin{abstract}

The axial charge of the triton is investigated using lattice quantum chromodynamics (QCD). Extending previous work at heavier quark masses, calculations are performed using three ensembles of gauge field configurations generated with quark masses corresponding to a pion mass of 450 MeV. Finite-volume energy levels for the triton, as well as for the deuteron and diproton systems, are extracted from analysis of correlation functions computed on these ensembles, and the corresponding  energies are extrapolated to infinite volume using finite-volume pionless effective field theory (FVEFT). It is found  with high likelihood that there is a compact bound state with the quantum numbers of the triton at these quark masses.  The axial current matrix elements are computed using background field techniques on one of the ensembles and FVEFT is again used to determine the axial charge of the proton and triton.
A simple quark mass extrapolation of these results and earlier calculations at heavier quark masses leads to a value of the ratio of the triton to proton axial charges at the physical quark masses of $g_A^{^{3}{\rm H}}/g_A^p=0.91${\raisebox{0.5ex}{\tiny$\substack{+0.07 \\ -0.09}$}}. This result is consistent with the ratio determined from experiment and prefers values less than unity (in which case the triton axial charge would be unmodified from that of the proton), thereby demonstrating that QCD can explain  the modification of the axial charge of the triton.
	 
\end{abstract}
\preprint{ICCUB-21-001}
\preprint{FERMILAB-PUB-21-026-T}
\preprint{MIT-CTP/5274}
\date\today
\maketitle

\section{Introduction}
\label{sec:intro}
The triton ($\hthree$) is the simplest nucleus that undergoes weak decay, and as such is an important system with which to investigate the Standard Model (SM) origins of nuclear physics.  While at the quark level the weak decay of the triton is  mediated by the weak interactions, it is the strong interactions, described by Quantum Chromodynamics (QCD), that dictate the embedding of quarks inside nuclei and thus are key to the nuclear decay rate. A notable feature of the $\beta$-decay of the triton and other nuclei is a reduction of the Gamow-Teller (isovector axial-vector) transition rate as compared to that of the neutron; this reduction scales approximately with the atomic number of the system, $A$, for medium mass nuclei \cite{Wilkinson:1973mgv,Brown:1985zz,Chou:1993zz,MartinezPinedo:1996vz} and can be described phenomenologically by an ad hoc reduction of the in-nucleus axial charge of the proton or by the introduction of two-nucleon interactions with the weak current \cite{Gysbers:2019uyb}. For the triton, this manifests as a suppression of the isovector axial matrix element; analysis of precision tritium decay experiments finds that the ratio of the axial charge of the triton to that of the proton is  $g_A^{{}^{3}{\rm H}}/g_A^p=0.9511(13)$ \cite{Baroni:2016xll}. This deviation from unity, and more generally, the difference of the axial charges of nuclei from that of the nucleon, is important phenomenologically and is a key aspect of nuclear physics to understand from the underlying SM. 
As well as probing our understanding of nuclear structure, study of tritium $\beta$-decay is a very promising avenue through which to improve constraints on neutrino masses \cite{Kraus:2004zw,Aseev:2011dq,Angrik:2005ep,Esfahani:2017dmu}. Careful comparison of decay measurements with theoretical predictions may also  reveal physics beyond that of the SM \cite{Gonzalez-Alonso:2018omy,Crivellin:2020lzu}.
Moreover, nuclear effects in  Gamow-Teller transitions are  important for understanding neutrinoful and neutrinoless double-$\beta$ decay matrix elements and thereby for sensitive searches for lepton number violation \cite{Engel:2016xgb}.

The half-life of tritium, $t_{1 / 2},$ is related to the Fermi ($\langle\mathbf{F}\rangle$) and Gamow-Teller ($\langle\mathbf{GT}\rangle$) reduced matrix
elements through 
\begin{equation}
 t_{1 / 2}= \frac{K / G_{V}^{2}}{\left(1+\delta_{R}\right) f_{V}}\frac{1}{\langle\mathbf{F}\rangle^{2}+f_{A} / f_{V} g_{A}^{2}\langle\mathbf{G} \mathbf{T}\rangle^{2}}\,,
\end{equation}
where the constants $K$, $G_V$, and $\delta_R$ are known precisely from theory or experiment~\cite{Simpson:1987zz}, $f_{A, V}$ denote known Fermi functions~\cite{Simpson:1987zz}, and $g_A\equiv g_A^p$ is the axial charge of the nucleon. From the Ademollo-Gatto theorem \cite{Ademollo:1964sr}, $\langle\mathbf{F}\rangle \sim 1$ and is only modified at second-order in isospin breaking and by electromagnetic corrections. 
The Gamow-Teller matrix element is less well determined and depends on the isovector axial current
\begin{equation}
\label{eq:isovecaxialcurrent}
    A_k^a= \bar{q} \gamma_{k} \gamma_{5} \tau^{a} q,
\end{equation}
where $q=(u,d)^T$ denotes a doublet of quark fields, $\gamma_\mu$ are Dirac matrices, and $\tau^a$ are Pauli matrices in flavor space.
In particular, the Gamow-Teller matrix element is defined from the zero-momentum projected current $\tilde A_k^a$ as
\begin{equation}
\left\langle\hethree,s\left|\tilde A_k^+ \right|\hthree,s^\prime\right\rangle= \bar{U}_s \gamma_{k} \gamma_{5} \tau^{+}U_{s^\prime}  g_{A}\langle\mathbf{GT}\rangle,
\label{eq:H3GTME}
\end{equation}
where $g_A^{{}^{3}{\rm H}}\equiv g_{A}\langle\mathbf{GT}\rangle$, $U_s$ is a relativistic spinor for the nucleus spin component $s\in\{+,-\}$, and $\tau^+=\tau_1+i\tau_2$. Eq.~\eqref{eq:H3GTME} is valid for zero electron mass and  vanishing electron momentum.

Because of the low energy scale of the $\beta$-decay process, determinations of the axial charges of hadronic and nuclear systems probe QCD in the non-perturbative regime; theoretical determinations must therefore  be undertaken using lattice QCD (LQCD) which is the only known systematically improvable tool for calculations in this regime. The axial charge of the proton has  been calculated using LQCD for many years following the first studies in Ref.~\cite{Fucito:1982ff}, see Ref.~\cite{Aoki:2019cca} for a recent summary of results. A first calculation of the axial charge of the triton was presented  in Ref.~\cite{Savage:2016kon}, albeit at an $SU(3)$-symmetric point with quark masses corresponding to  a pion  mass $m_\pi=806$ MeV. This  work  extends Ref.~\cite{Savage:2016kon} with calculations at quark masses that are considerably closer to their physical values, corresponding to $m_\pi=450$ MeV and a kaon mass  $m_K= 595$ MeV \cite{Orginos:2015aya}. At these quark masses, the infinite volume extrapolated Gamow-Teller matrix element is determined to be $\langle\mathbf{GT}\rangle_{L\to\infty}=0.938(41)$. Combined with the earlier work,  extrapolations of this result to the physical quark masses leads to $\langle\mathbf{GT}\rangle=g_A^{\hthree}/g_A^p=0.91${\raisebox{0.5ex}{\tiny$\substack{+0.07 \\ -0.09}$}}. These are the main results of this work and show that the phenomenological modification of the axial charge of the triton can be reproduced directly from QCD. 

\section{Lattice QCD Details}
\label{sec:lattice}

The lattice QCD calculations presented in this work make use of isotropic gluon configurations generated with a L\"uscher-Weisz \cite{Luscher:1984xn} gauge action and $N_f=2+1$ flavors of fermions implemented using a tadpole-improved clover quark action \cite{Sheikholeslami:1985ij}. All  ensembles are generated using a gauge coupling of $\beta=6.1$ and with degenerate up and down quark masses corresponding to a pion mass of $m_\pi=450$ MeV and a strange quark mass that corresponds to a kaon mass of $m_K=595$ MeV. 
Performing scale setting using upsilon spectroscopy extrapolated to the physical quark masses results in a lattice spacing of $a=0.1167(16)$ fm.
These configurations have previously been used to study two-baryon interactions \cite{Orginos:2015aya,Illa:2020nsi} and further details are presented in Ref.~\cite{Orginos:2015aya}. Three different lattice volumes are used, as shown in Table \ref{tab:latt}.
\begin{table}[!t]
	\begin{ruledtabular}
\begin{tabular}{c|cccccccc}
      Label & $L/a$ & $T/a$ &  $L$ [fm] & $T$
      [fm] & $m_\pi L$ & $m_\pi T$ & $N_{\rm cfg}$ & $N_{\rm meas}$\\
\hline
      E24 & 24 & 64  & 2.80 &  7.47 &   6.390 & 17.04 & 2124 & $5.6 \times 10^5$\\
      E32 & 32 & 96   & 3.73 & 11.2 & 8.514 & 25.54  & 2850 & $2.8 \times 10^5$ \\	
      E48 & 48 & 96  & 5.60 & 11.2 & 12.78 & 25.49 & 980 & $4.7 \times 10^4$ 
\end{tabular}
 \end{ruledtabular}
	\caption{Details of the ensembles of gauge field configurations used in this calculation, and of the measurements performed on them. In all cases,  the gauge coupling is $\beta=6.1$, the tadpole-improved Sheikholeslami-Wohlert parameter is $c_{\rm SW}=1.24931$ \cite{Sheikholeslami:1985ij}, and the bare light and strange quark masses in lattice units are $a m_{ud}^{(\rm bare)}=-0.2800$ and  $a m_{s}^{(\rm bare)}=-0.2450$, respectively. Each ensemble consists of $N_{\rm cfg}$ configurations on which a total of $N_{\rm meas}$ measurements were made (slightly different numbers of sources were used on each configuration).
		\label{tab:latt}}
\end{table}

\section{Spectroscopy and infinite volume extrapolation at $m_\pi\sim 450$ MeV}
\label{sec:spectro}

In order to determine  the ground-state energy of the triton and $^3$He, which are degenerate for the isospin-symmetric quark masses used in this calculation, two-point correlation functions projected to zero three-momentum are constructed using the methodology introduced in Refs.~\cite{Detmold:2012eu,Beane:2012vq}. The correlation functions are %
\begin{equation}
C_{h,s}(t) = \sum_{\bf x}\Gamma^{(s)}_{\beta\alpha} \left\langle  \chi_\alpha^{(h)\prime}({\bf x} ,t) \chi_\beta^{(h)\dagger}({\bf 0},0) \right\rangle
\quad {\rm for}\ s\in\{+,-\},
\label{eq:interpdef}
\end{equation}
where $\Gamma^{(\pm)}=\frac{1}{4}(1+\gamma_4)(1\pm i \gamma_1 \gamma_2)$ is a projector onto the given positive energy spin component and $\chi_\alpha^{(h)\prime}$ and  $\chi_\alpha^{(h)}$ are interpolating operators with the quantum numbers of the hadron $h\in\{p,\hthree\}$. For the triton, the interpolating operator is built from three color-singlet nucleons that are independently projected to zero three-momentum in Eq.~\eqref{eq:interpdef}.  
The quark propagators used to construct the correlation functions are computed using APE smeared \cite{Falcioni:1984ei} sources and point or APE smeared sinks; the resulting correlation functions are referred to as `SP' and `SS' respectively.\footnote{For the $L=\{24, 32, 48\}$ ensembles, smearing parameters of $N_{\rm smear} = \{80, 35, 50\}$ steps of smearing with Gaussian smearing parameters $\rho=\{3.5, 4.7, 3.5\}$ were used.
}
An advantage of using local multi-hadron sources in this manner is that they can be efficiently computed for light nuclei such as the triton using baryon block algorithms whose cost scales linearly with the spatial volume~\cite{Detmold:2012eu}.
A disadvantage of this approach is that since source and sink operators are distinct, it is not possible to build a variational basis of operators that would explicitly account for excited-state contamination from unbound multi-nucleon states that have small energy separations to the ground state for large volumes.
Similar issues arise in the two-nucleon sector, and Ref.~\cite{Illa:2020nsi} discusses  a number of consistency checks that have been applied to two-nucleon results using multiple interpolating operators on the same ensemble as used here. Given tensions between two-nucleon ground-state energy results  with $m_\pi \sim 800$ MeV obtained using products of zero-momentum baryon sources \cite{Francis:2018qch,Horz:2020zvv} compared with results obtained using local or approximately local two-baryon sources~\cite{Beane:2012vq,Wagman:2017tmp}, it will be important to pursue variational studies of both the two-nucleon and three-nucleon sectors, including operators overlapping with bound and unbound states, in the future. However, multi-nucleon variational studies require a large set of interpolating operators whose correlation functions are significantly more costly to compute than those calculated here and are deferred to future work.
While here the focus is on the triton and the nucleon, other single hadron and two-baryon systems have been studied using the same approach as applied here, on the same ensembles of gauge field configurations, as discussed in Refs.~\cite{Orginos:2015aya,Illa:2020nsi}.

The ground-state triton energy and its difference from the mass of three non-interacting nucleons are extracted in each volume from analysis of the time dependence of $C_{{}^3{\rm H},\pm}(t)$ and $C_{p,\pm}(t)$ using  the same fitting methodology as applied and detailed in Refs.~\cite{Beane:2020ycc,Illa:2020nsi}. For completeness, the approach is summarized here. Provided that the temporal separation of the source and sink, $t$, is larger than the extent of the lattice action, and small compared with the temporal extent of the lattice geometry ($t\ll T$), the correlation functions are given by a sum of exponentials whose exponents are determined by the energies of the eigenstates of the given quantum numbers: 
\begin{equation}
\label{eq:expsum}
    C^{{\cal S}{\cal S}^\prime}_{h,\pm}(t)=\sum_{n=0}^{N_{\rm ex}} {Z_n^{(h)\cal S}Z_n^{(h){\cal S}^\prime\ast}} \exp(-E_n^{(h)}t)
    \quad {\rm for}\ \{{\cal S},{\cal S}^\prime\} \in\{{\rm S},{\rm P}\},
\end{equation}
where $N_{\rm ex}$ excited states contribute to the sum, $Z_n^{(h)\text{S/P}}$ denotes the overlap factor of the corresponding  interpolating operators onto the $n$th energy eigenstate, and thermal effects arising from the finite temporal extent of the lattice geometry have been neglected. Fits of the correlation functions to Eq.~\eqref{eq:expsum} determine the ground state energies $M_h=E_0^{(h)}$ among the fit parameters; while correlation functions for different propagator smearings have different overlap factors, the energies are common and are thus fit simultaneously.
To quantify the systematic uncertainties arising from the choice of source-sink separations, $t$, included in the fit, and of the truncation of the sum in Eq.~\eqref{eq:expsum}, 200 fit windows are sampled at random from the set of all choices of contiguous time-ranges longer than $t_{\rm plat}=6$ and with final times less than $t_{\rm max}$ (defined by the point at which the noise-to-signal ratio exceeds $tol_{\rm noise}=0.25$ for the given correlation function).
For each fit range, the Akaike Information Criterion~\cite{1100705} (AIC) is used to perform model selection ({\it i.e.}, fix the number of exponential terms  in the sum above). The truncation is set as the largest $N_{\rm ex}$ for which the change in AIC is $\Delta \text{AIC} < - 0.5 N_{\rm dof}$, where $N_{\rm dof}$ is the number of degrees of freedom of the fit. 
In each case, combined correlated fits are performed to average correlation functions as well as to $N_{\rm boot}=200$ bootstrap resamplings of the correlation functions using covariance matrices computed using optimal shrinkage \cite{stein1956,LEDOIT2004365,Rinaldi:2019thf} and using  variable projection (VarPro)  \cite{varpro0,varpro1} to eliminate overlap factors. All fits with a $\chi^2/N_{\rm dof}< tol_{\chi^2}=2$ are included in a set of `accepted' fits (accepted fits must also pass tests  that  the fit results are a) invariant to the choice of the minimizer that is used to within $tol_{\rm solver} = 10^{-5}$, b) agree within a prescribed tolerance of $tol_{\rm corr} = 5\sigma$ with uncorrelated fits, and c) agree within a prescribed tolerance of $tol_{\rm median} = 2\sigma$ to the median bootstrap result, as in Refs.~\cite{Illa:2020nsi,Beane:2020ycc}). The final value and uncertainties for the energy are then computed from the results of all $N_{\rm success}$ accepted fits using a weighted average. The central value and statistical and systematic uncertainties are computed as
\begin{equation}
      \overline{E}_0 = \sum_{f=1}^{N_{\rm success}} w^f E_0^f, \hspace{1cm}
       \delta_{\rm stat} \overline{E}_0 = \delta E_0^{f_{\rm max}} ,\hspace{1cm}
      \delta_{\rm sys} \overline{E}_0^2 = \sum_{f=1}^{N_{\rm success}} w^f \left( E_0^f - \overline{E}_0 \right)^2,
\label{eq:weightedave}
\end{equation}
where $E_0^f$ denotes the fit result from a given fit labelled by $f$, and $w_f$ is the associated weight.
For each fit range, the statistical uncertainties $\delta E_0^f$ are computed using 67\% confidence intervals determined from the results of the $N_{\rm boot}$ resampled correlation function fits described above; the total statistical uncertainty is defined as the statistical uncertainty of the fit $f_{\rm max}$ with maximum weight $w_{f_{\rm max}}$.
The statistical and systematic uncertainties are combined in quadrature to give a total uncertainty $\delta \overline{E}_0 = \sqrt{\delta_{\rm stat} \overline{E}_0^2 + \delta_{\rm sys} \overline{E}_0^2}$. 
Since each successful fit provides a statistically unbiased estimate of the ground-state energy, the relative weights $w_f$ of each fit in the weighted average can be chosen arbitrarily (in the limit of large statistics)~\cite{Rinaldi:2019thf}. Here, as in Refs.~\cite{Rinaldi:2019thf,Illa:2020nsi,Beane:2020ycc}, the weights are set as
\begin{equation}
   \begin{split}
      w^f &= \frac{p_f \left( \delta E_0^f\right)^{-2} }{ \sum_{f^\prime = 1}^{N_{\rm success}} p_{f^\prime} \left( \delta E_0^{f^\prime} \right)^{-2}  },
   \end{split}\label{eq:weights}
\end{equation}
where $p_f = \Gamma(N_{\rm dof}/2, \chi_f^2/2)/\Gamma(N_{\rm dof}/2)$ is the $p$-value of fit $f$ assuming $\chi^2$-distributed goodness-of-fit parameters.

\begin{figure}[!t]
	\includegraphics[width=0.45\columnwidth]{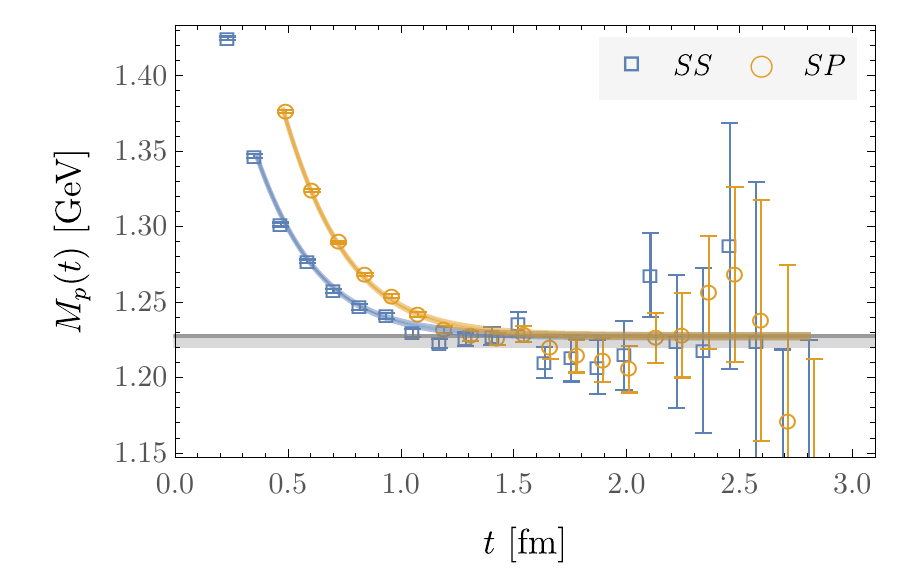}
	\includegraphics[width=0.45\columnwidth]{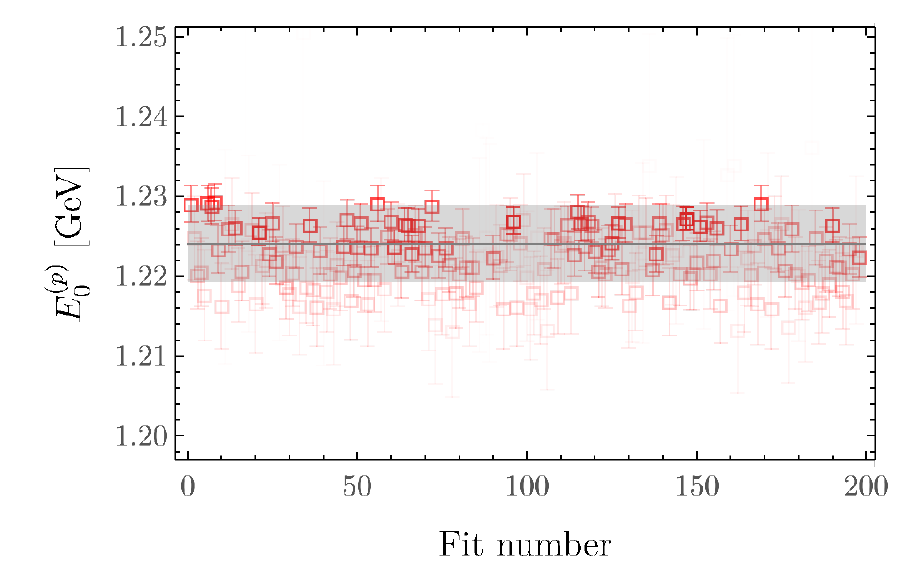}
	\includegraphics[width=0.45\columnwidth]{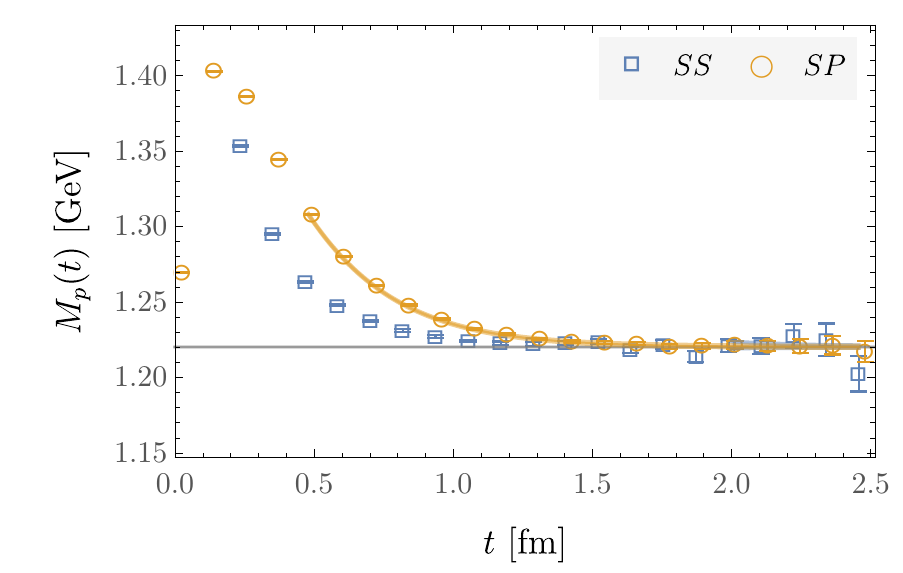}
	\includegraphics[width=0.45\columnwidth]{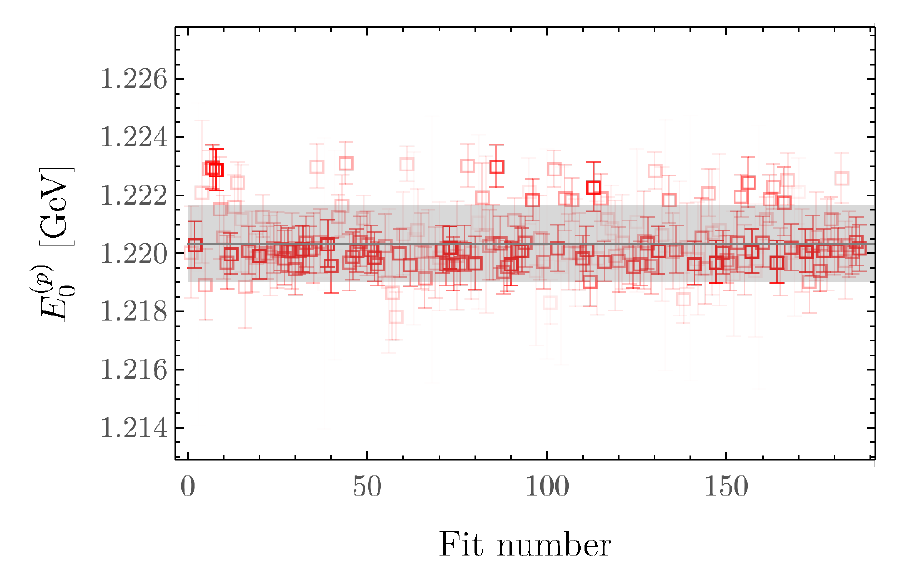}
	\includegraphics[width=0.45\columnwidth]{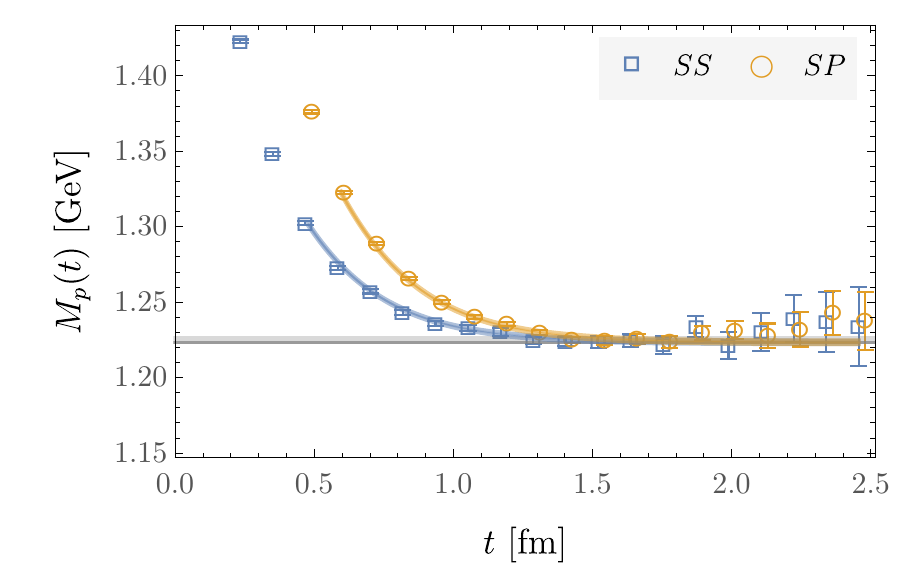}
	\includegraphics[width=0.45\columnwidth]{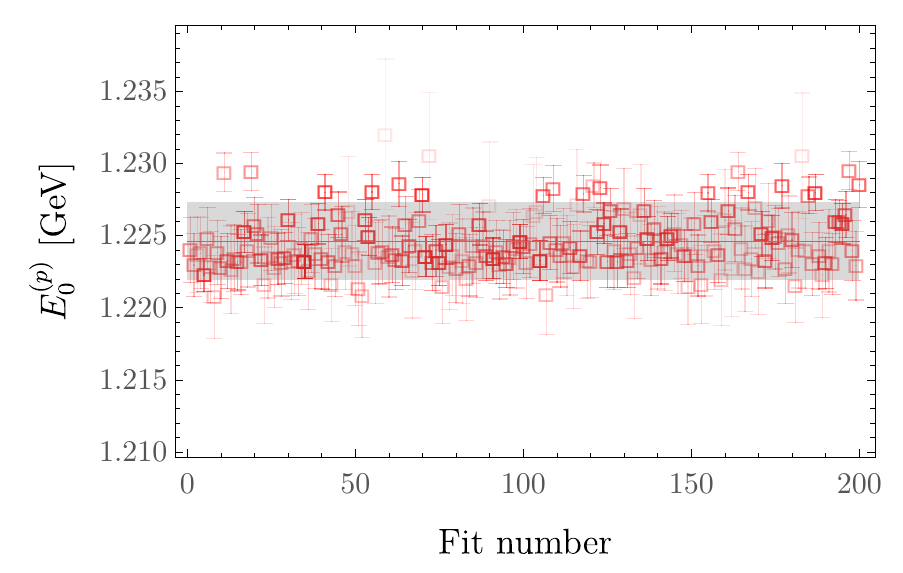}
	\caption{\label{fig:proteffmass} Left column: effective mass functions (Eq.~\eqref{eq:effmass}) for the proton correlation function on the E24 (top), E32 (center), and E48 (bottom) ensembles. The orange circles and blue squares show effective masses constructed from SP and SS correlation functions respectively, and the corresponding orange and blue curves display the highest-weight fit contributing to the weighted average of Eq.~\eqref{eq:weightedave}; the light (dark) grey bands show the mass extracted from the weighted average (single highest-weight fit). Right column: masses extracted from each successful fit to the correlators. The opacity of each point is set by the contribution of the fit to the weighted average. The horizontal line and grey band in each figure correspond to the final result for the mass and its total uncertainty.}
\end{figure}
\begin{figure}[!t]
	\includegraphics[width=0.45\columnwidth]{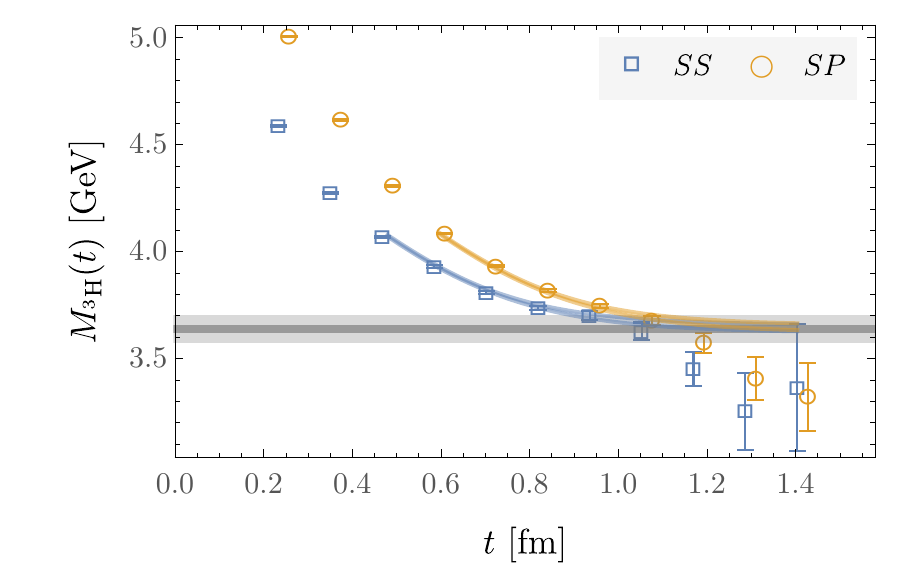}
	\includegraphics[width=0.45\columnwidth]{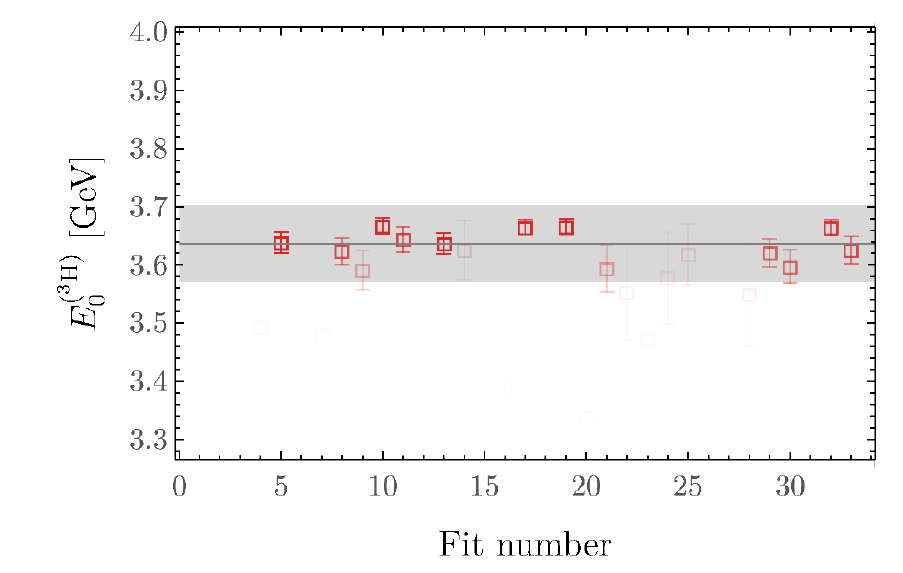}
	\includegraphics[width=0.45\columnwidth]{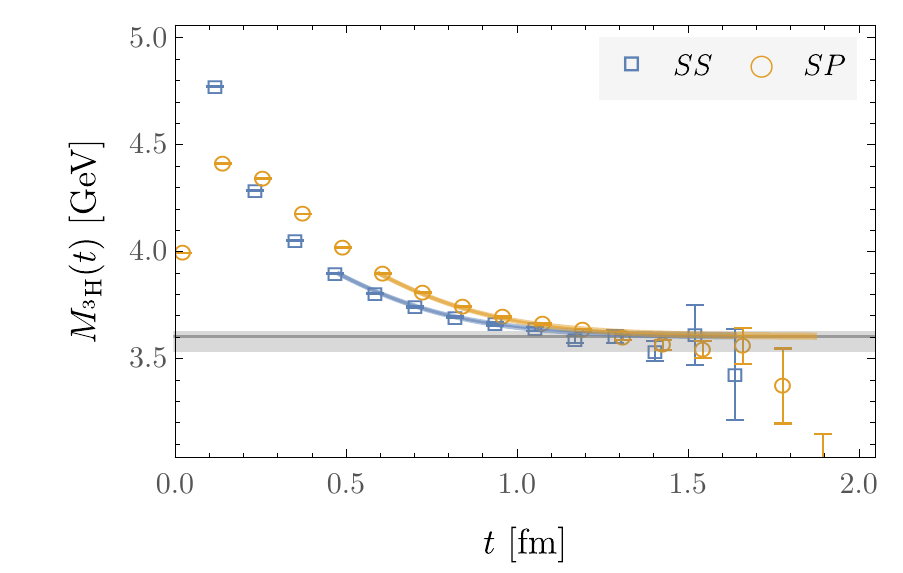}
	\includegraphics[width=0.45\columnwidth]{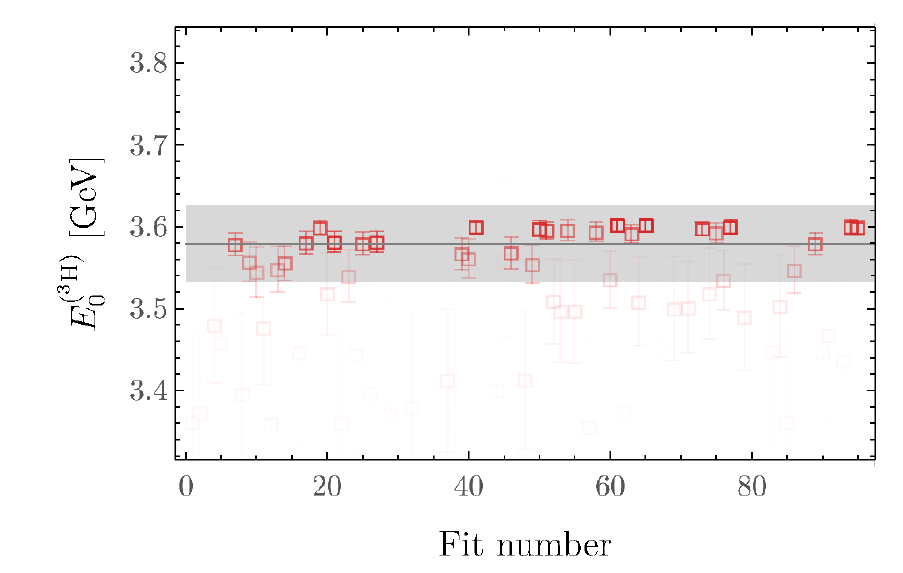}
	\includegraphics[width=0.45\columnwidth]{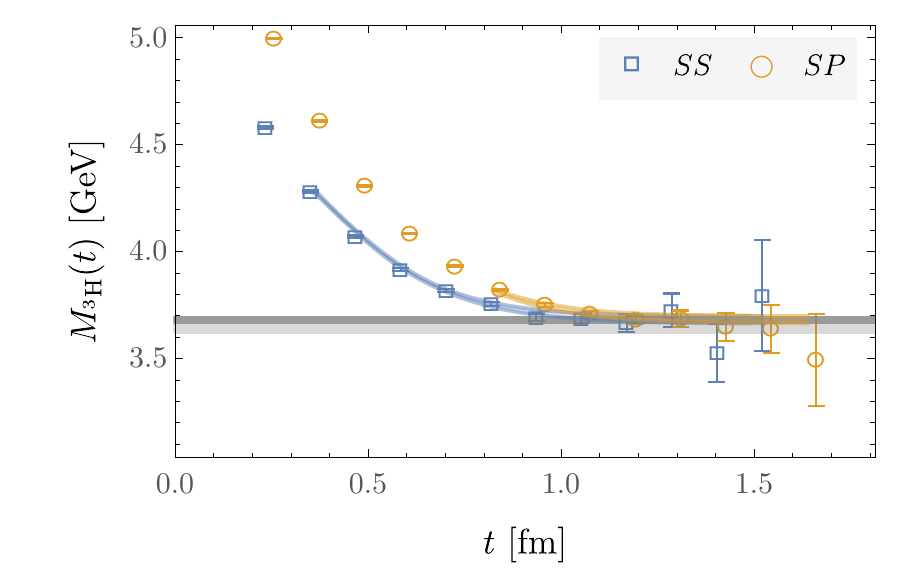}
	\includegraphics[width=0.45\columnwidth]{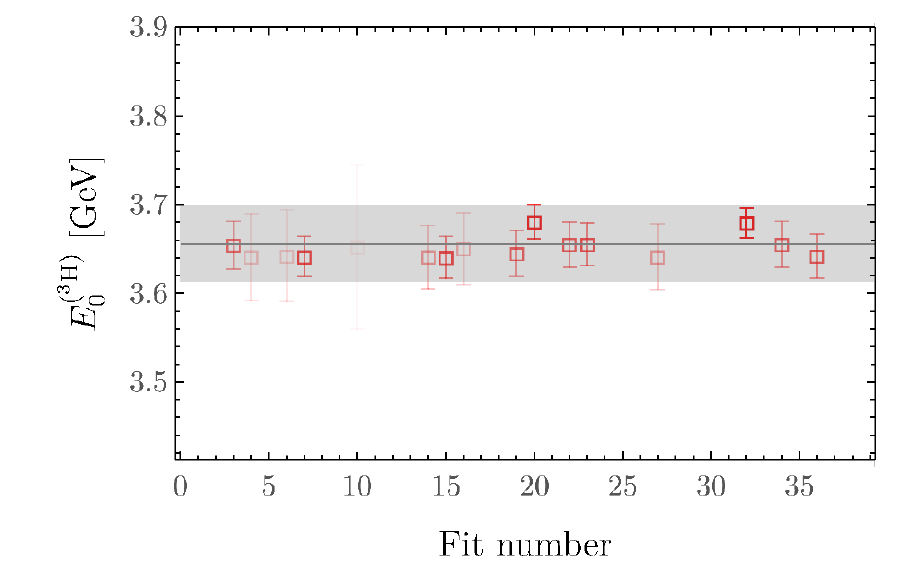}
	\caption{\label{fig:tritoneffmass} Effective mass functions and fit summaries for the fits of the triton correlation function on the E24 (top), E32 (center), and E48 (bottom) ensembles. All features are as in Fig.~\ref{fig:proteffmass}.}
\end{figure}

The resulting fitted masses are summarised for each volume in Table~\ref{tab:results} and the fits are shown graphically for the proton in Fig.~\ref{fig:proteffmass} and for the triton in  Fig.~\ref{fig:tritoneffmass}.
In each figure, the effective mass functions  
\begin{equation}\label{eq:effmass}
    aM^{\mathcal{SS'}}_h(t) = \ln \left( \frac{\sum_{s=\pm}C^{\mathcal{SS'}}_{h,s}(t)}{\sum_{s=\pm}C^{\mathcal{SS'}}_{h,s}(t+a)} \right),
\end{equation} 
are shown for the SS and SP  data along with the functional form of best fits to the correlation functions and the final result for $E_0^{(h)}$.
The difference between the triton mass and three times the proton mass, $\Delta E_{{}^3{\rm H}}= E_0^{({}^3{\rm H})}-3E_0^{(p)}$, is determined from the correlated differences of the fitted energies computed under bootstrap.
The results for this quantity are also presented in Table~\ref{tab:results}.
\begin{table}[!t]
	\begin{ruledtabular}
		\begin{tabular}{cccc}
			Ensemble	& $a E_0^{(p)}$      &  $a E_0^{({}^3{\rm H})}$         &  $a \Delta E_{{}^3{\rm H}}$       \\ \hline
			E24 & $0.7258(29)$ & $2.157(40)$ & $-0.024(32)$\\
			E32 & $0.7236(8)$  & $2.122(28)$ & $-0.054(24)$\\
			E48 & $0.7261(16)$ & $2.168(26)$ & $-0.012(14)$\\
		\end{tabular}
	\end{ruledtabular}
	\caption{Fit results for the proton and triton ground-state energies, as well as the difference $\Delta E_{{}^3{\rm H}}= E_0^{({}^3{\rm H})}-3E_0^{(p)}$, determined in lattice units on each ensemble. 
	\label{tab:results}}
\end{table}

\begin{figure}[!t]
\subfigure[]{
	\includegraphics[width=0.6\columnwidth]{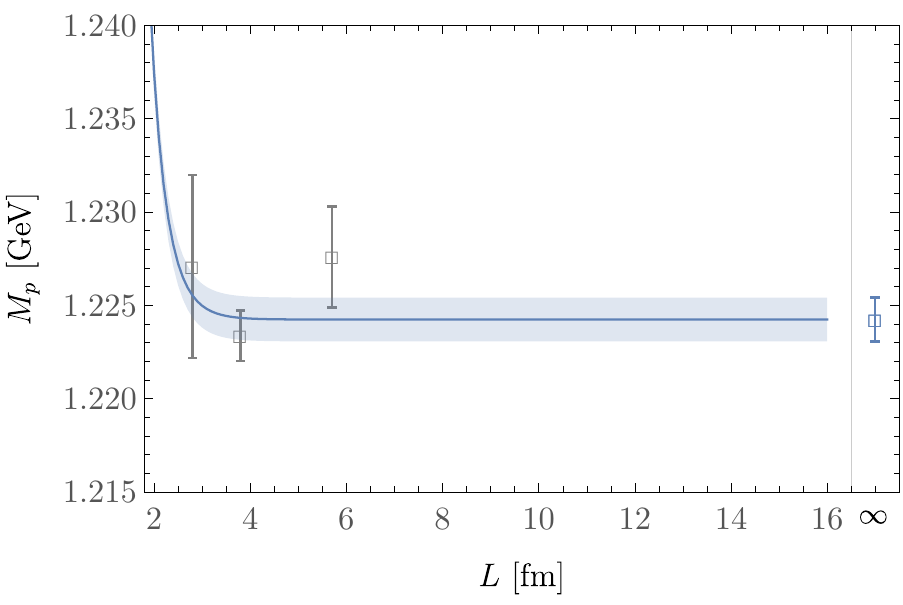}
}
\subfigure[]{
	\includegraphics[width=0.6\columnwidth]{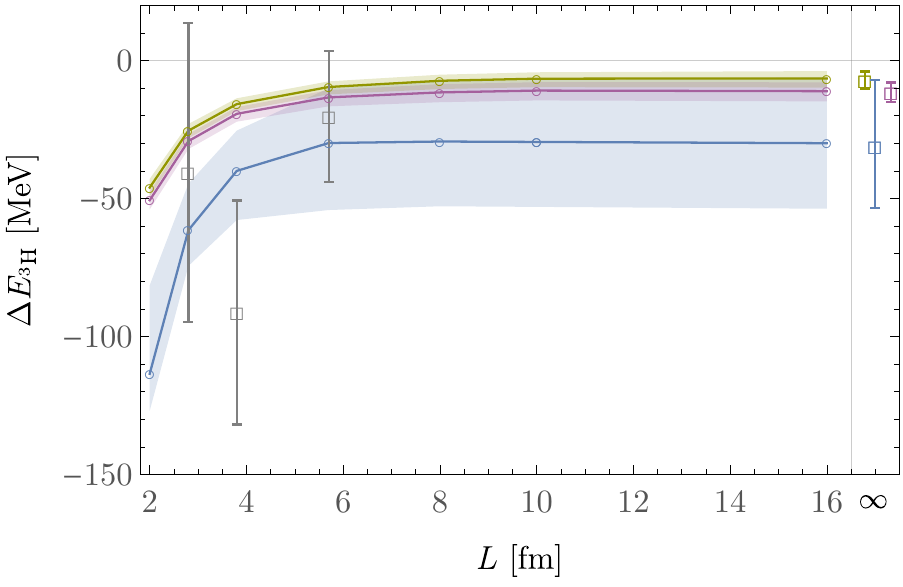}
}
\caption{\label{fig:MpvsL} 
(a) Mass of  the proton as a function of the spatial extent of the lattice geometry, along with the fit to this dependence via Eq.~\eqref{eq:MNvsL}.
(b)	The binding energy of the triton as a function of the spatial extent of the lattice geometry. Grey squares correspond to the binding energies determined in the LQCD calculations, while the blue curve shows the finite-volume dependence in pionless EFT fitted to the LQCD data. The blue point at the right indicates the infinite volume extrapolation of the binding energy. For comparison, the binding energies of the $pp$ (purple) and deuteron (green) systems obtained via an analogous FVEFT study are also shown.}
\end{figure}

In the limit in which $m_\pi L$ is sufficiently large and the pion mass is sufficiently small that the volume dependence of the proton mass is described by $p$-regime chiral perturbation theory ($\chi$PT), that dependence takes the form \cite{AliKhan:2003ack,Beane:2004tw}
\begin{eqnarray}
M_p(L)- M_p^\infty =-\frac{3 g_{A}^{2}}{8 \pi^{2} f_\pi^{2}} \mathcal{K}(0,L)-\frac{g_{\Delta N}^{2}}{3 \pi^{2} f_\pi^{2}} \mathcal{K}(\Delta,L)\,,
\label{eq:MNvsL}
\end{eqnarray}
where $f_\pi$ is the chiral limit pion decay constant and 
\begin{eqnarray}
\mathcal{K}(\Delta,L) \equiv \int_{0}^{\infty} d \lambda \beta_{\Delta} \sum_{\vec{n} \neq \vec{0}}\left[(L|\vec{n}|)^{-1} K_{1}\left(L \beta_{\Delta}|\vec{n}|\right)-\beta_{\Delta} K_{0}\left(L \beta_{\Delta}|\vec{n}|\right)\right]\,.
\label{eq:KFV}
\end{eqnarray}
Here, $L$ denotes the spatial extent of the lattice geometry, $K_n$ are modified Bessel functions,  $\beta_{\Delta}^{2} \equiv \lambda^{2}+2 \lambda \Delta+m_{\pi}^{2}$, $\Delta$ denotes the mass splitting between the nucleon and the $\Delta$ baryon, and $g_{\Delta N}$ is the nucleon-$\Delta$ transition axial coupling.  The sum in Eq.~\eqref{eq:KFV} is over integer triplets excluding $\vec{n}=(0,0,0)$, and  for $\Delta=0$ the asymptotic behavior of Eq.~(\ref{eq:MNvsL}) is $\sim e^{-m_\pi L}/L$. While $m_\pi=450$~MeV is not in the regime where chiral perturbation theory shows rapid convergence for baryon properties, it has been found to effectively describe the  volume-dependence of baryon masses in this mass regime~\cite{Beane:2011pc}. With the physical values of $g_A=1.26$, $g_{\Delta N}=1.4$,   $f_\pi=132$ MeV and $\Delta=300$ MeV, Eq.~(\ref{eq:MNvsL}) is used to fit to the infinite volume proton mass from the masses determined on the three volumes. This fit, displayed in Figure \ref{fig:MpvsL}, results in an infinite volume mass of $M_p^\infty=1.2242(12)$~GeV.\footnote{This is in agreement with the value $M_p^\infty=1.226(12)$~GeV found in Ref.~\cite{Orginos:2015aya,Illa:2020nsi} from the same ensembles. Note that fits using the values of the axial charges and $\Delta$--nucleon mass splitting determined at the quark masses used in the calculations give very similar extrapolated values.}
 
Figure \ref{fig:MpvsL} also shows the difference between the triton energy and the three-nucleon threshold for each of the three volumes. Unlike for the proton, the form of the volume dependence of the triton energy is not known analytically, however a numerical description can be found by matching to finite-volume effective field theory (FVEFT). The procedure of matching pionless EFT to LQCD results for the binding energies of light nuclei using the stochastic variational method has been demonstrated in Ref.~\cite{Eliyahu:2019nkz} using LQCD results with $m_\pi= 806$~MeV. The same procedure as detailed in that work is followed for the data presented here, and further details of the FVEFT approach used here are provided in Appendix~\ref{app:FVEFT}.\footnote{As well as the EFT approach, recent generalizations of the  quantization conditions derived by L\"uscher for two particles relate finite-volume three-particle energies to the infinite-volume three-particle scattering amplitude and constrain bound states when present, see Ref.~\cite{Briceno:2017max} for a recent review.}  
The infinite volume extrapolation leads to an energy shift $\Delta E_{{}^3{\rm H}}^{\infty}=-32(23)$ MeV. The FVEFT formalism is compatible with both scattering states and bound systems and the extracted energy suggests the state is not consistent with three scattering nucleons. This leaves the possibility that it is a compact three-body bound state or either a deuteron--neutron or dineutron-proton scattering state, as the binding energies of the deuteron and dineutron are $7.2(3.2)$ and $11.6(3.6)$ MeV from the FVEFT extrapolation respectively (illustrated in Fig.~\ref{fig:2bodyfits} in Appendix ~\ref{app:FVEFT};  note also that these results are consistent, within 1 standard deviation, with those obtained from this data via L\"uscher's method in Ref.~\cite{Orginos:2015aya,Illa:2020nsi}).  While the latter cases of 2+1 body systems can not be ruled out, there is a strong preference (80\% likelihood, using the most conservative two-body binding energies) that the state is a compact three-body system. In what follows, this is assumed to be the case and the state is referred to as the `triton'.

\begin{figure}[!t]
	\includegraphics[width=0.6\columnwidth]{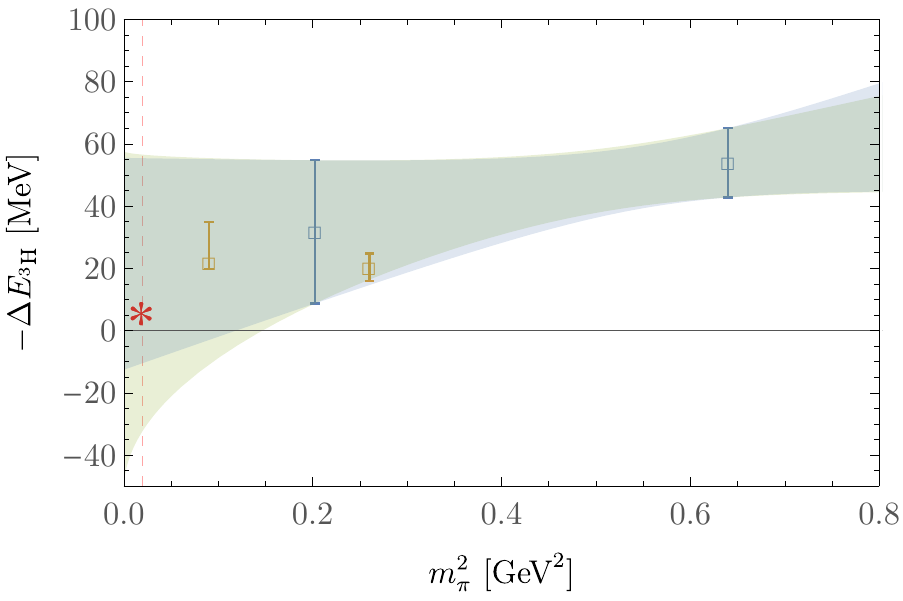}
	\caption{\label{fig:binding} 
	Binding energy of the triton as a function of the pion mass. The current result is shown in blue, along with the result obtained in a previous calculation \cite{Beane:2012vq} using the same action at $m_\pi=806$ MeV. The results of Refs.~\cite{Yamazaki:2012hi,Yamazaki:2015asa} are shown in orange. The physical value of the binding is shown as a red star at the physical pion mass, as indicated by the vertical line. Linear and quadratic fits in $m_\pi$ to the blue data are shown as the green and blue bands respectively.}
\end{figure}

In Fig.~\ref{fig:binding}, the resulting binding energy of the triton is compared to the results of other calculations, including that of Ref.~\cite{Beane:2012vq} using the same action but  heavier quark masses corresponding to $m_\pi=806$ MeV. The extracted binding energy is compatible with those of other calculations at nearby quark masses \cite{Yamazaki:2012hi,Yamazaki:2015asa}, although no effort is made to take into account the differences between the lattice actions or scale-setting schemes that are used. Naive extrapolations of the current result and that from Ref.~\cite{Beane:2012vq} that are linear or quadratic in $m_\pi$ are consistent with the experimental value for the binding energy of the triton, albeit with large uncertainties. The strong evidence for binding at the other masses shown in the figure, and the assumption of smooth behavior under variation in the pion mass  provides additional support for the conclusion that the triton is a compact three-body system at these quark masses.

\section{Gamow-Teller matrix element for tritium $\beta$-decay}
\label{sec:GTME}

To extract the axial charge of the triton, and hence the Gamow-Teller contribution to tritium $\beta$-decay, LQCD calculations of matrix elements of the isovector axial current in the triton and proton are performed using the E32 ensemble. Only a single ensemble is used due to computational cost and for technical simplicity, the flavor-diagonal matrix element $\langle \hthree | \tilde A_3^3 | \hthree\rangle$ is studied and is related by isospin symmetry to the tritium $\beta$-decay matrix element. 
The resulting values are matched to FVEFT to determine the relevant low energy constants (LECs), which are then used to predict the infinite-volume matrix element.

To compute the finite-volume matrix elements, the extended propagator technique discussed in detail in Refs.~\cite{Savage:2016kon,Shanahan:2017bgi,Tiburzi:2017iux,Bouchard:2016heu,Davoudi:2020ngi} is used. This requires extraction of hadronic and nuclear correlation functions at a range of values of an applied constant axial field that couples to up and down quarks separately. Extended quark propagators are defined as
\begin{equation}
S_{\lambda_{f}}^{(f)}(x, y)=S^{(f)}(x, y)+\lambda_f \int d^4 z S^{(f)}(x, z) \gamma_3\gamma_5 S^{(f)}(z, y), 
\end{equation}
where $S^{(f)}(x, y)$ is a quark propagator of flavor $f\in\{u,d\}$ and $\lambda_{f}$ is the strength of the applied field for the given flavor. These quantities are calculated for five values of the external field strength $\lambda_{u,d}\in \{0,\pm0.05,\pm0.1\}$ in lattice units. Two-point correlation functions $C_{h,s}^{(\lambda_u,\lambda_d)}(t)$ are constructed from these extended propagators using the same contraction methods as for the zero-field correlation functions discussed in the previous section. For clarity, the smearing labels SS/SP are suppressed in this section. The two-point correlation functions are polynomials in $\lambda_f$ of order $n_f$, the number of quark propagators of flavor $f$ in the correlation function. With computations at $n_f$ different field strength values, the terms in the polynomial can thus be extracted uniquely~\cite{Tiburzi:2017iux} and are labelled as $\left.C_{h,s}^{(\lambda_f,0)}\right|_{{\cal O}(\lambda_f^k)}$ and $\left.C_{h,s}^{(0,\lambda_f)}\right|_{{\cal O}(\lambda_f^k)}$ for $k\in \{0,\ldots,n_f\}$. It is straightforward to show~\cite{Savage:2016kon,Tiburzi:2017iux} through insertions of complete sets of states that an effective isovector axial charge function, which asymptotes to the desired bare axial charge in the corresponding hadronic or nuclear state at large temporal separations, can be defined as
\begin{equation}
  \label{eq:effcharge}
  g_A^{h}(t) = R_{h}(t+a) - R_{h}(t) \underset{t\rightarrow\infty}{\longrightarrow} g_A^{h(0)} + {\cal O}(e^{-\delta t}),
\end{equation} 
where the $(0)$ superscript denotes that the charge is not renormalized, $\delta$ is the energy difference between the ground state and the first excited state with the quantum numbers of the state $h$, and
\begin{align}
\label{eq:Rratio}
R_{h}(t)=\sum_{s=\pm} \eta_s \frac{\left.C_{h,s}^{(\lambda_u,0)}(t)\right|_{{\cal O}(\lambda_u)} - \left.C_{h,s}^{(0,\lambda_d)}(t)\right|_{{{\cal O}(\lambda_d)}}}{2a\,C_{h,s}^{(0,0)}(t)}
\end{align} 
for  $\eta_\pm=\mp1$.

The effective charges in Eq.~\eqref{eq:effcharge} are constructed from sums of ratios of two-point functions whose time dependence is each of the generic exponential form in Eq.~(\ref{eq:expsum}).  The axial charges can therefore be isolated by fits to the time-dependence of the effective charge functions. These fits are performed using an extension of the  fit range sampling and excited-state model selection procedure discussed above to background field three-point functions   (see Ref.~\cite{Detmold:2020snb} for further details).
The spectral representations for the $\mathcal{O}(\lambda)$ correlation functions appearing in Eq.~\eqref{eq:Rratio} can be constructed as\footnote{The correlation function is defined in Euclidean spacetime, and the sum over $\tau$ extends only over the temporal range between the source and the sink because of the scalar isoscalar nature of the vacuum (exponentially small contributions that are suppressed by the mass of the lightest axial-vector meson are ignored).}
\begin{equation}
\begin{split}
    \left. C_{h,s}^{(\lambda_u,0)}(t)\right|_{{\cal O}(\lambda_u)} -  \left. C_{h,s}^{(0,\lambda_d)}(t)\right|_{{\cal O}(\lambda_d)} &= a  \eta_s \sum_{\tau/a = 0}^{t/a} \sum_{n,m} {Z_n^{(h)}Z_m^{(h)\prime\ast}}  e^{-E_n^{(h)} (t-\tau)} e^{-E_m^{(h)} \tau} g_A^{h(m,n)(0)},
    \end{split}
    \label{eq:BGspec}
\end{equation}
where the prime on the second $Z$-factor is included to denote that, although smearings are suppressed in this section, the overlap factors differ at the source and sink, and the bare (transition) charge $g_A^{h(m,n)(0)}$ is defined from the corresponding matrix element as
\begin{eqnarray}
    \mbraket{h,n,s}{\tilde A_3^3}{h,m,s^\prime} = \bar{U}_{n,s} \gamma_3\gamma_5 \tau_3 U_{m,s^\prime} g_A^{h(m,n)(0)},
\end{eqnarray}
where $\ket{h,n,s}$ denotes states with the quantum numbers of $h$ and $U_{m,s}$ denotes the spinor for state $m$ with spin $s$.
This can be used to derive a spectral representation for $g_A^h(t)$:
\begin{align}\nonumber
        g_A^h(t) &= \sum_n{Z_n^{(h)} Z_n^{(h)\prime\ast}} g_A^{h(n,n)(0)}  \left(  \frac{ (t/a+1) e^{-E_n^{(h)} (t+a)}}{\sum_k {Z_k^{(h)} Z_k^{(h)\prime\ast}}  e^{-E_k^{(h)} (t+a)}} - \frac{(t/a)\, e^{-E_n^{(h)} t}}{\sum_k {Z_k^{(h)} Z_k^{(h)\prime\ast}} e^{-E_k^{(h)} t}} \right)   \\&+ \mathcal{O}(a),
\label{eq:Rspec}
\end{align}
where the $\mathcal{O}(a)$ contribution is detailed in Ref.~\cite{Detmold:2020snb} and depends on excited-state transition matrix elements as well as combinations of overlap factors not determined from fits of two-point functions to Eq.~\eqref{eq:expsum}.
Notably, the $\mathcal{O}(a)$ term is absent for a single-state correlation function model.
Multi-state fits have been performed both with and without these $\mathcal{O}(a)$ terms and the AIC is used to determine whether the $\mathcal{O}(a)$ terms should be included in the fit.
For both the triton and proton, this AIC test prefers the fit without $\mathcal{O}(a)$ terms for all fit range choices.
Combined fits of $\frac{1}{2}\sum_{s=\pm}C_{h,s}(t)$ to Eq.~\eqref{eq:expsum} and $g_A^h(t)$ to Eq.~\eqref{eq:Rspec} without $\mathcal{O}(a)$ terms using both SP/SS interpolating operator combinations are therefore used to determine $E_n^{(h)}$, the product ${Z_n^{(h)} Z_n^{(h)\prime\ast}}$, and $g_A^{h(n,n)}$.
For each fitting interval, $\ln (E_n^{(h)})$ and $g_A^{h(0)} = g_A^{h(n,n)(0)}$ are used as nonlinear optimizer parameters with $Z_n^{(h)}$ obtained from $C_{h,s}(t)$ using VarPro as above and $g_A^{h(n,n)(0)}$ subsequently obtained from $g_A^h(t)$ using VarPro.
Statistical uncertainties on the ground-state matrix elements for each fit are obtained using bootstrap confidence intervals, and a weighted average performed analogously to the two-point function case described above is used to determine the final ground-state matrix element values and statistical plus fitting systematic uncertainties.
Results for the ground-state matrix elements $g_A^{h{(0)}}$ obtained using this fitting procedure for both one- and three-nucleon systems are shown in Table~\ref{tab:effchargeesults}.

\begin{figure}[!t]
	\includegraphics[width=0.49\columnwidth]{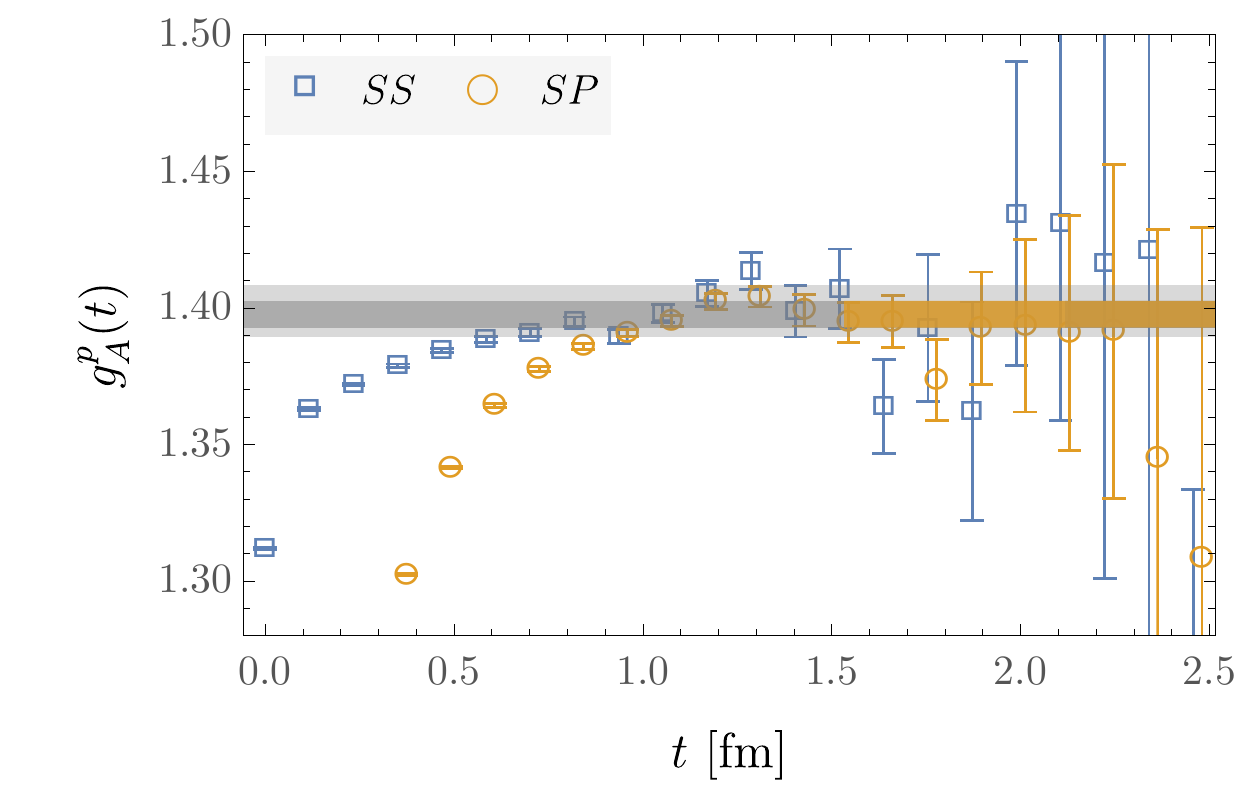}
	\includegraphics[width=0.49\columnwidth]{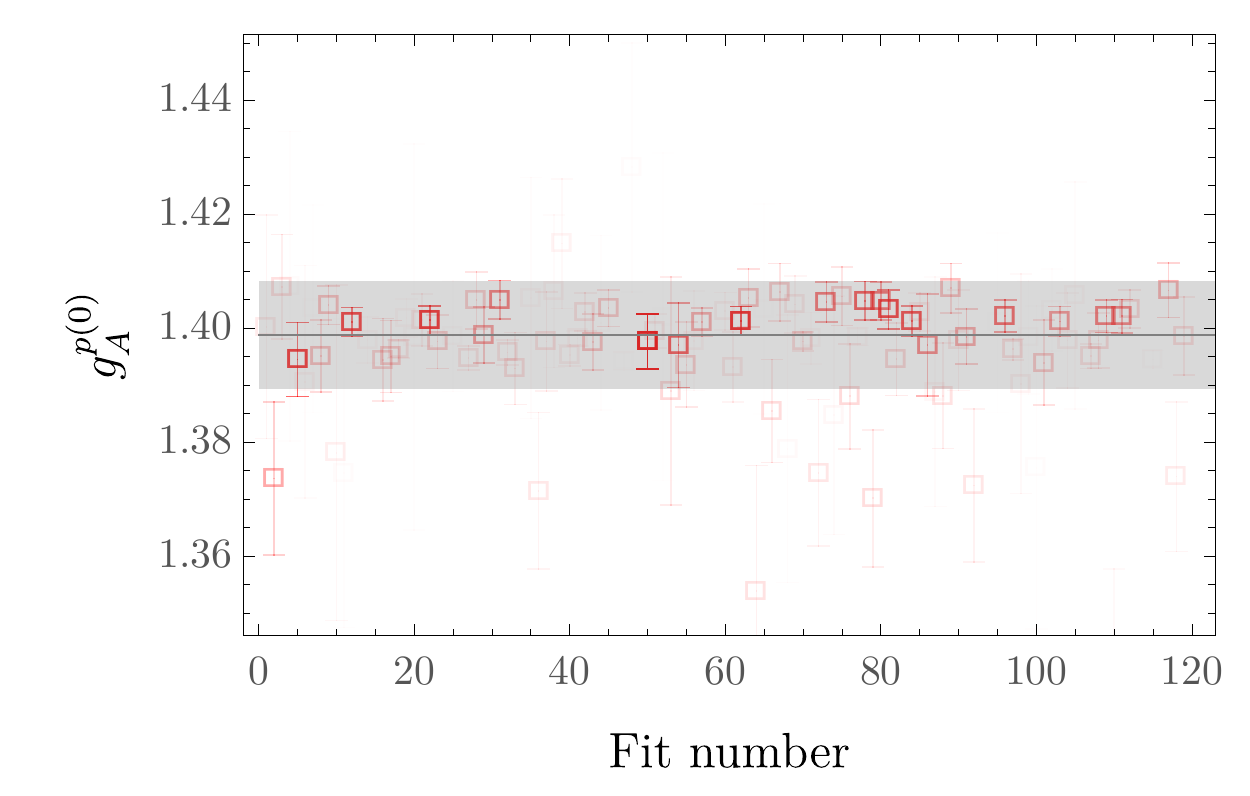}
	\includegraphics[width=0.49\columnwidth]{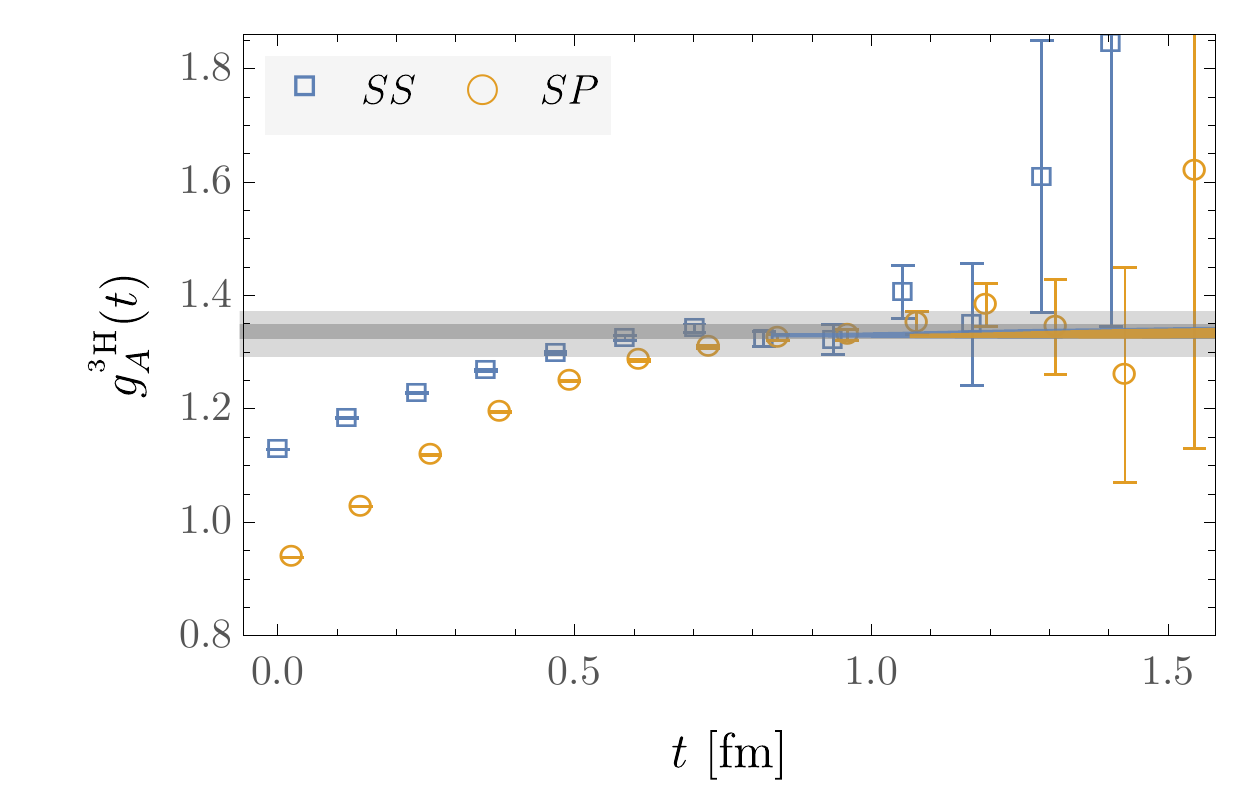}
	\includegraphics[width=0.49\columnwidth]{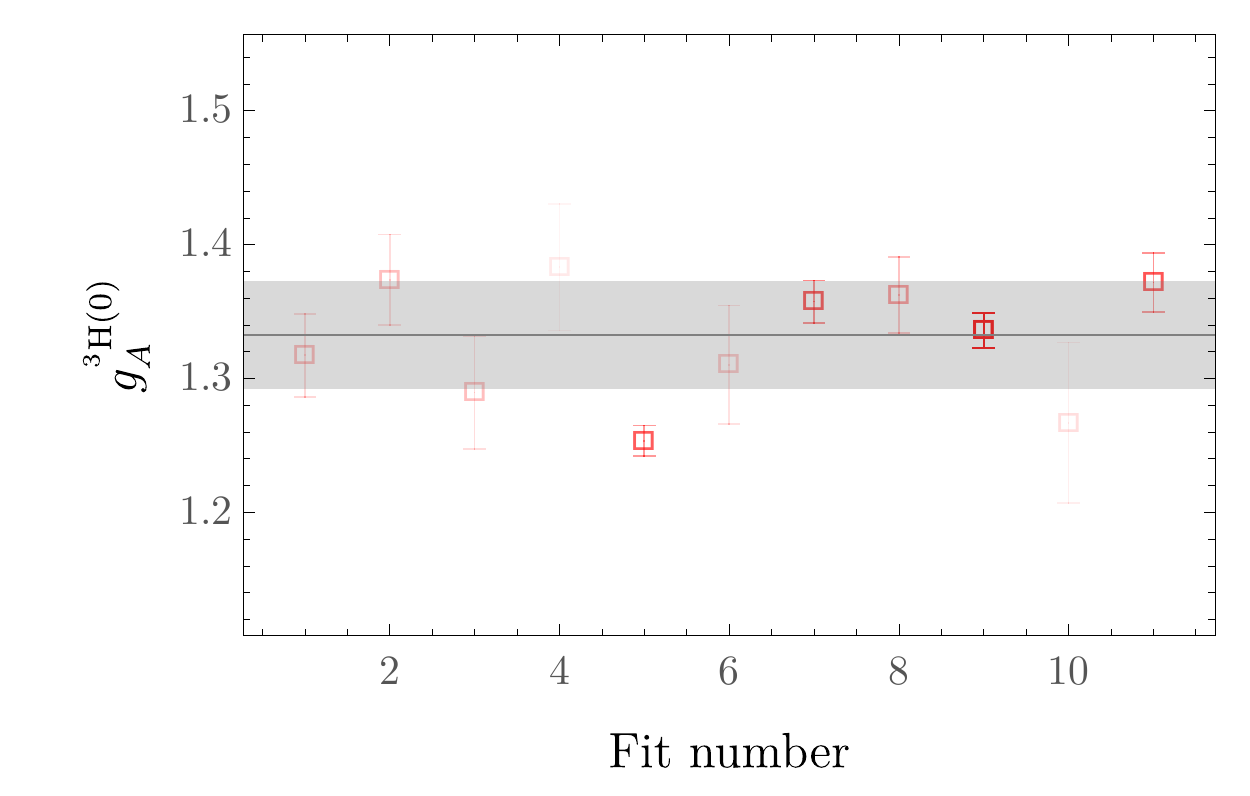}
	\caption{\label{fig:effcharge} The left panels show the proton and triton bare effective matrix element functions, $g_A^p(t)$ and $g_A^{{}^3{\rm H}}(t)$, respectively, calculated on the E32 ensemble. In each figure, the orange (blue) data correspond to the SP (SS) calculations and the corresponding shaded bands illustrate the highest weight fit. The light (dark) shaded gray bands denote the extracted values of the matrix elements arising from the combined analysis (single highest-weight fit). The right panels show the values obtained for all the successful fits to different time ranges, with the opacity determined by the contribution of the fit in the final weighted average. The  horizontal line and gray band in the right panels show the final central value and uncertainty.
	}
\end{figure}
\begin{figure}[!t]
	\includegraphics[width=0.49\columnwidth]{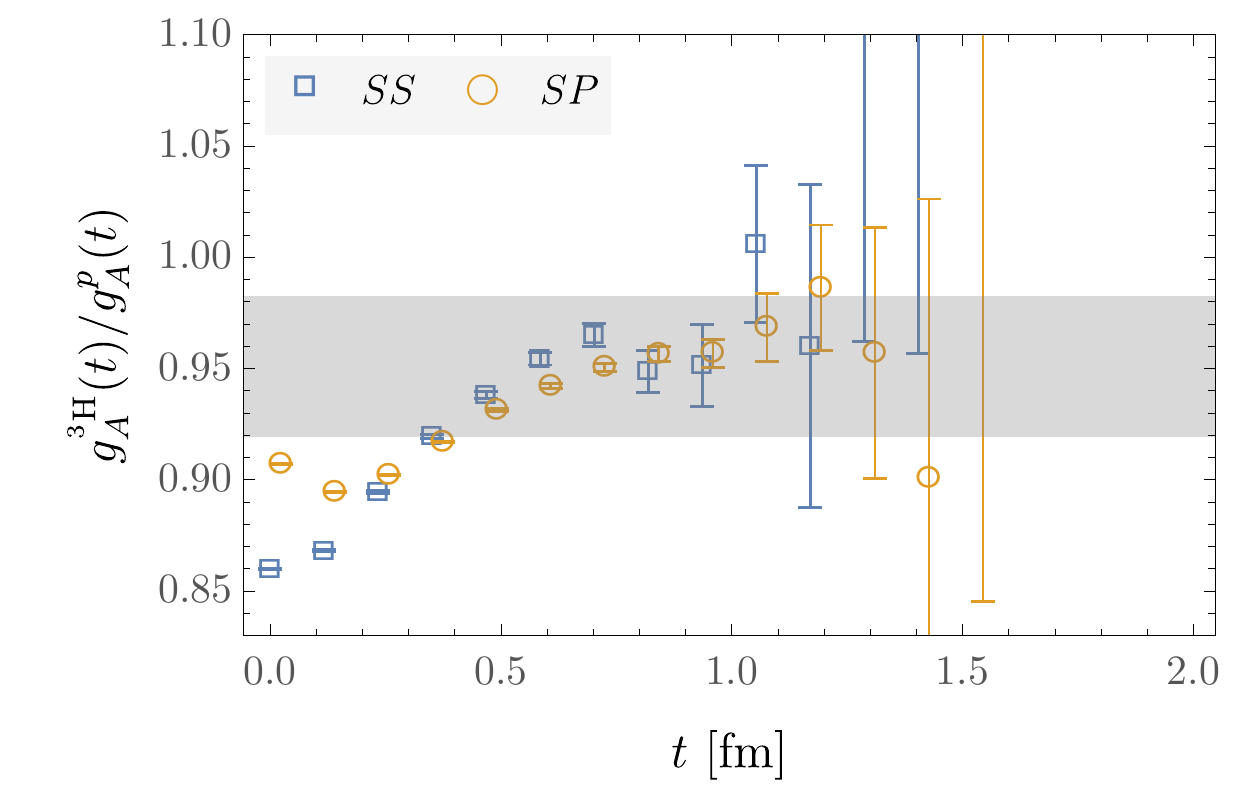}
	\caption{\label{fig:effchargeratio} The effective ratio function $g_A^{^{3}{\rm H}}(t)/g_A^p(t)$, constructed through Eq.~(\ref{eq:effcharge}), which asymptotes to  $g_A^{\hthree}/g_A^p$ and is independent of axial current renormalization.  The orange (blue) data show the SP (SS) correlation functions on the E32 ensemble and the shaded band indicates the extracted value of the matrix element (not a fit to the displayed data).
	}
\end{figure}

The quantities $g_A^p(t)$ and $g_A^{{}^3{\rm H}}(t)$ constructed on ensemble E32 for both SS and SP source--sink pairs are shown in Fig.~\ref{fig:effcharge}. Also shown are the values of the bare matrix elements determined from fits to the time-dependence of these functions as discussed above. Table \ref{tab:effchargeesults} displays the extracted bare couplings $g_A^{h(0)}$ as well as renormalized couplings $g_A^h = Z_A g_A^{h(0)}$ obtained by multiplying by the appropriate axial current renormalization constant, $Z_A=0.8623(01)(71)$ determined in Ref.~\cite{Yoon:2016jzj}.
The ratio $g_A^{{}^3{\rm H}}(t)/g_A^p(t)$, which at large times asymptotes to the GT reduced matrix element $\langle\mathbf{GT}\rangle=g_A^{{}^3{\rm H}}/g_A^{p}$, and is independent of the renormalization of the axial current, is shown in Fig.~\ref{fig:effchargeratio} along with the value of $\langle\mathbf{GT}\rangle$ that is extracted from the fits to the individual axial couplings.
\begin{table}[!t]
	\begin{ruledtabular}
		\begin{tabular}{cccc}
			h	&  $  g_A^{h(0)} $ &  $g_A^h$  &  $g_A^h/g_A^p$  \\ \hline
			$p$ & $1.399(9)$ & $1.230(19)$  & --
						\\
			${}^3{\rm H}$ & $1.332(40)$ & $1.171(39)$  & $0.953(30)$
		\end{tabular}
	\end{ruledtabular}
	\caption{Fit results for the bare, $g_A^{h(0)}$, and renormalized, $g_A^{h}$, axial charges of the proton and triton as well as the corresponding ratio to the axial charge of the proton. The statistical uncertainties and fitting systematic uncertainties are combined in quadrature. The fitting systematic uncertainties account for variation in the choice of fit ranges and multi-state fit models used to obtain $g_A^{h (0)}$ and the resulting correlated ratio of ground-state matrix elements $g_A^h/g_A^p$ as described in the main text.  
	\label{tab:effchargeesults}}
\end{table}

The FV three-body matrix element in Table \ref{tab:effchargeesults} can be used to constrain the leading two-nucleon axial current counterterm of pionless effective field theory in the finite volume. To do so, the approach developed in Ref.~\cite{WP} is followed, whereby EFT wavefunctions, determined variationally and matched to the LQCD spectrum computed in the E32 volume, are used to evaluate the FVEFT matrix elements of the EFT current:
\begin{equation}
\begin{aligned}
{\cal A}_{i,a}=& \frac{g_{A}}{2} N^{\dagger} \tau_{a} \sigma_{i} N 
+L_{1, A}\left[\left(N^{T} P_{i} N\right)^{\dagger}\left(N^{T} \bar{P}_{a} N\right)+\text{h.c.}\right] 
+\ldots,
\end{aligned}
\label{eq:axialcurrent}
\end{equation}
where $g_A$ is the single-nucleon axial coupling, and the projectors
\begin{equation}
\begin{aligned}
P_{i} & \equiv \frac{1}{\sqrt{8}} \sigma_{2} \sigma_{i} \tau_{2}, \qquad
\overline{P}_{a} & \equiv \frac{1}{\sqrt{8}} \sigma_{2} \tau_{2} \tau_{a},
\end{aligned}
\label{eq:projectors}
\end{equation}
form spin-triplet and spin-singlet two-nucleon states, respectively.
The two-body couterterm $L_{1,A}$ is regulator-dependent, and in this work a Gaussian regulator scheme is used with a scale $\Lambda$ as discussed in Appendix \ref{app:FVEFT}.

\begin{figure}[!t]
	\subfigure[]{
	\includegraphics[width=0.485\columnwidth]{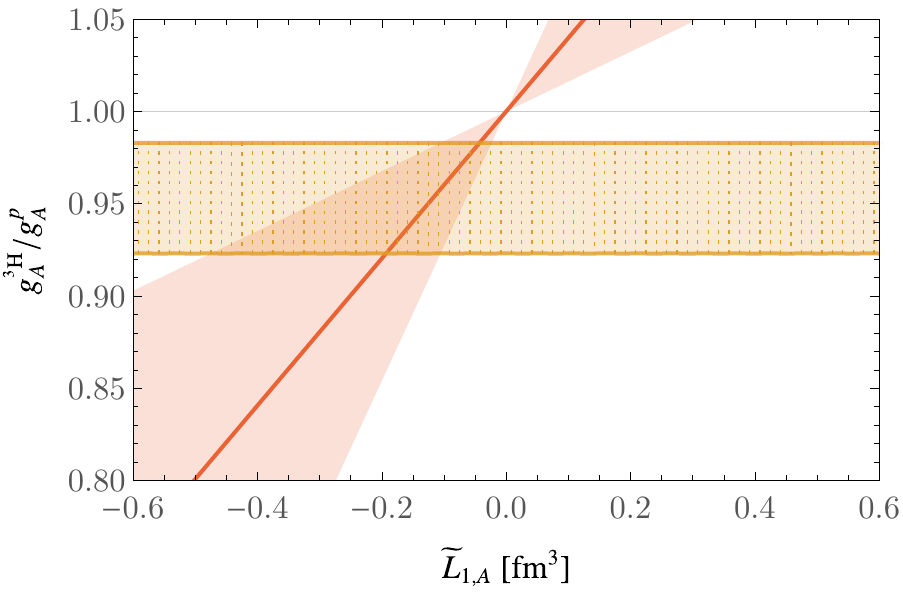}}
	\subfigure[]{
	\includegraphics[width=0.485\columnwidth]{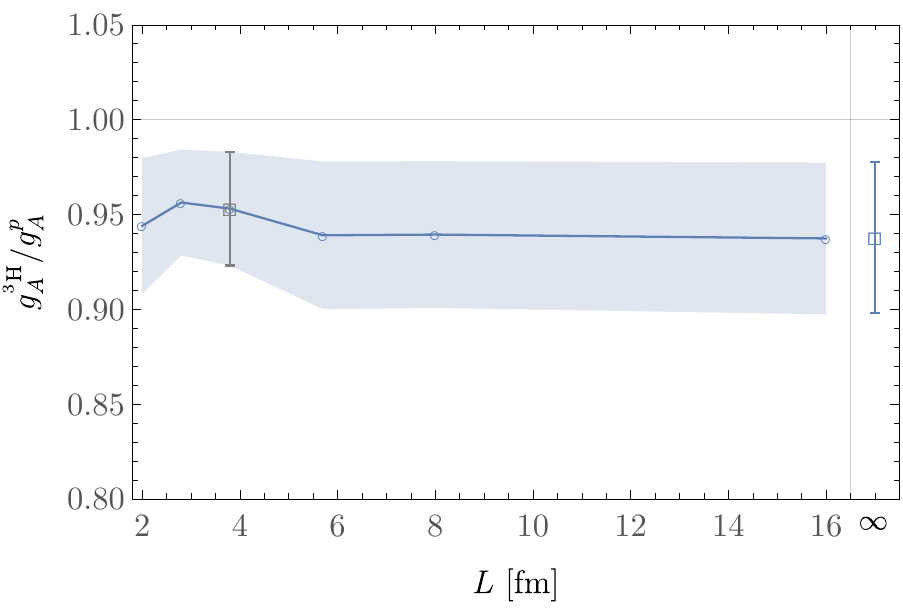}}
	\caption{\label{fig:effchargeatiovsL} (a) Determination of the LEC ratio $\widetilde{L}_{1,A}$ by matching the matrix element defined in Eq.~\eqref{eq:EFTdude}, computed in the optimized FVEFT wavefunction for the E32 volume, to the LQCD result in the E32 volume, which is depicted as the horizontal band. (b) The FVEFT extrapolation of the ratio of triton to proton axial charges to infinite volume. }
\end{figure}

The ratio of the triton to proton matrix elements in FVEFT  is given by 
\begin{equation}
    \frac{2}{g_A}\frac{\langle \Psi_{\hthree}, s=+ | {\cal A}_{3,3} | \Psi_{\hthree}, s=+\rangle}{ \langle \Psi_{\hthree}, s=+ | \Psi_{\hthree}, s=+\rangle}
    =  \left( 1 + \frac{L_{1,A}}{3g_A} 
    h_{\hthree}(\Lambda,L) \right)\,,
    \label{eq:EFTdude}
\end{equation}
where $| \Psi_{\hthree}, s\rangle$ is the wavefunction for the spin-$s$ $\hthree$ state and $ h_{\hthree}(\Lambda,L)$ is the spatial expectation value of a regulated form of ${\cal A}_{3,3}$ in the triton spatial wavefunction as detailed in Appendix \ref{app:FVEFT}.
The  ratio of LECs $\widetilde{L}_{1,A}\equiv L_{1,A}/g_A$ is determined by demanding that  Eq.~\eqref{eq:EFTdude} for the E32 volume reproduces the LQCD ratio of  axial charges for $^3\text{H}$ and for the proton computed on that volume.
This value of $\widetilde{L}_{1,A}$ is then used to compute the axial current matrix element in variationally-optimised triton wavefunctions for different volumes, including the infinite volume wave function that allows the infinite-volume matrix element to be determined (more details on this procedure are given in Appendix \ref{app:FVEFT}). While the counterterm $L_{1,A}$ is scheme and scale-dependent, the triton axial charge for any volume is scale-independent. Figure \ref{fig:effchargeatiovsL} shows the result of this matching procedure and the volume dependence of the ratio of triton to proton axial matrix elements. As expected from the deep binding of this system, the volume effects are small. The resulting infinite-volume GT matrix element is $\langle\mathbf{GT}\rangle_{L=\infty}=g_A^{{}^3{\rm H}}/g_A^{p}=0.938(41)$. 

\section{Discussion}
\label{sec:discussion}

\begin{figure}[!t]
	\includegraphics[width=0.6\columnwidth]{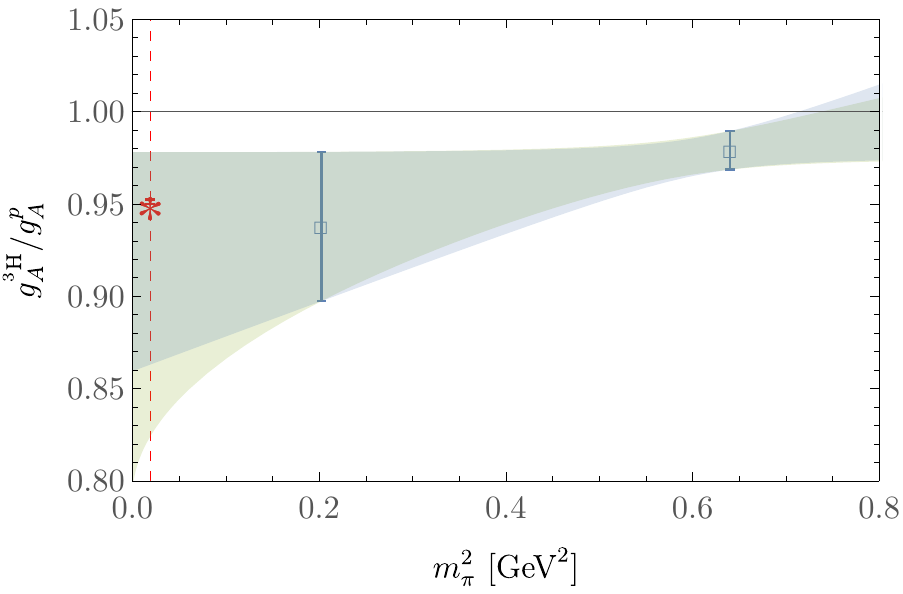}
	\caption{\label{fig:effchargeatio} Ratio of the axial charge of tritium to that of the single nucleon as a function of the pion mass. The result from this work and that of Ref.~\cite{Savage:2016kon} are shown as the blue points while the physical value \cite{Baroni:2016xll} is shown in red at the physical pion mass (indicated by the vertical line). }
\end{figure}

In order to connect to the physical quark masses, the result for the ratio of triton and proton axial charges described in the last section can be combined with the previous determination of $g_A^{^{3}{\rm H}}/g_A^p= 0.979(03)(10)$ in Ref.~\cite{Savage:2016kon} at the $SU(3)$-symmetric point (where $m_\pi=m_K=806$ MeV) using the same action and scale setting procedure.
Tritium $\beta$-decay has been investigated in $\chi$PT in Ref.~\cite{Baroni:2018fdn} (see Refs.~\cite{De-Leon:2016wyu,De-Leon:2019dqq} for related work in pionless effective field theory) and  so the mass dependence of the ratio is in principle known. However, the quark masses in the calculation of Ref.~\cite{Savage:2016kon}, and potentially in the current work, are beyond the regime of applicability of $\chi$PT. Additionally, for the three-body system of the triton, the effective field theory results are determined numerically rather than analytically. 
At present, the above discussion, and the paucity of LQCD data, motivates extrapolation of the axial current matrix element ratio with the phenomenological forms of linear and quadratic dependence on the pion mass.
The calculated GT matrix elements and both of these extrapolations are shown in Fig.~\ref{fig:effchargeatio}; the two fits result in values of 0.90(8) and 0.92(6), respectively. 
Given the model dependence of the forms used in this extrapolation, the envelope of the extrapolated uncertainties is taken as the extrapolated result, leading to $g_A^{^{3}{\rm H}}/g_A^p=0.91${\raisebox{0.5ex}{\tiny$\substack{+0.07 \\ -0.09}$}}.

While extrapolated to infinite volume and the physical quark masses, these results are determined at a single lattice spacing and QED and isospin breaking effects are absent and the uncertainties from these systematic effects can as yet only be estimated. Lattice spacing effects are expected to contribute to the matrix elements at ${\cal O}(a \Lambda_{\rm QCD})$, however it is likely that there will be partial cancellations in these effects between the proton and triton axial charges and a full evaluation of this uncertainty will require further calculations.
The leading QED effects cancel in the ratio of triton to proton axial charges and isospin breaking effects in $g_A^p$ have been estimated as $\alt 1$ \% \cite{Jin:1996nj,Kaiser:2001yc} and are assumed to be similarly small for the triton. 
Exponentially-suppressed FV effects due to virtual pions are neglected in the pionless FVEFT formalism that has been used. However, for the volumes and masses used in this work and Ref.~\cite{Savage:2016kon}, $e^{-m_\pi L}\sim 10^{-4}$ so these effects are expected to be negligible.

The axial charge of the triton is thus extracted from lattice QCD for the first time in this work. 
The extrapolated coupling ratio is in agreement with the phenomenological value $g_A^{^{3}{\rm H}}/g_A^p=0.9511(13)$ \cite{Baroni:2016xll} and thus this calculation demonstrates the QCD origins of nuclear effects in the GT tritium $\beta$-decay matrix element. In the future, the two-body pionless EFT currents that are determined using these methods will also allow for calculations of the GT matrix elements in larger nuclei and a more comprehensive investigation of the QCD origin of the phenomenological quenching of the axial charge.

\acknowledgements
We are grateful to S. R. Beane, Z. Davoudi, K. Orginos, M. J. Savage, and B. Tiburzi for extensive discussions and collaboration in the early stages of this work. This research used resources of the Oak Ridge Leadership Computing Facility at the Oak Ridge National Laboratory, which is supported by the Office of Science of the U.S. Department of Energy under Contract number DE-AC05-00OR22725, as well as facilities of the USQCD Collaboration, which are funded by the Office of Science of the U.S. Department of Energy, and the resources of the National Energy Research Scientific Computing Center (NERSC), a U.S. Department of Energy Office of Science User Facility operated under Contract No. DE-AC02-05CH11231. The authors thankfully acknowledge the computer resources at MareNostrum and the technical support provided by Barcelona Supercomputing Center (RES-FI-2019-2-0032 and RES-FI-2019-3-0024). The Chroma~\cite{Edwards:2004sx}, QLua \cite{qlua},  QUDA \cite{Clark:2009wm,Babich:2011np,Clark:2016rdz}, QDP-JIT \cite{6877336} and QPhiX~\cite{10.1007/978-3-319-46079-6_30} software libraries were used in this work. WD and PES are supported in part by the U.S.~Department of Energy, Office of Science, Office of Nuclear Physics under grant Contract Number DE-SC0011090. WD is also supported within the framework of the TMD Topical Collaboration of the U.S.~Department of Energy, Office of Science, Office of Nuclear Physics, and  by the SciDAC4 award DE-SC0018121. PES is additionally supported by the National Science Foundation under EAGER grant 2035015, by the U.S. DOE Early Career Award DE-SC0021006, by a NEC research award, and by the Carl G and Shirley Sontheimer Research Fund. 
MI is supported by the Universitat de Barcelona through the scholarship APIF. MI and AP acknowledge support from the Spanish Ministerio de Economía y Competitividad (MINECO) under the project MDM-2014-0369 of ICCUB (Unidad de Excelencia “Mar\'ia de Maeztu”), from the European FEDER funds under the contract FIS2017-87534-P and by the EU STRONG-2020 project under the program H2020-INFRAIA-2018-1, grant agreement No. 824093. MI acknowledges the Massachusetts Institute of Technology for hospitality and partial support during preliminary stages of this work.
This manuscript has been authored by Fermi Research Alliance, LLC under Contract No. DE-AC02-07CH11359 with the U.S. Department of Energy, Office of Science, Office of High Energy Physics.  This work is supported by the U.S. Department of Energy, Office of Science, Office of Nuclear Physics under contract DE-AC05-06OR23177.
The authors acknowledge support from the U.S. Department of Energy, Office of Science, Office of Advanced Scientific Computing Research and Office of Nuclear Physics, Scientific Discovery through Advanced Computing (SciDAC) program, and of the U.S. Department of Energy Exascale Computing Project.
The authors thank Robert Edwards, B\'{a}lint Jo\'{o}, Kostas Orginos, and the NPLQCD collaboration for generating and allowing access to the ensembles used in this study.

\appendix
\section{FVEFT}
\label{app:FVEFT}

Following Ref.~\cite{WP}, the stochastic variational method is used to connect the finite volume axial matrix element to its infinite volume limit in pionless EFT. 
The pionless EFT Lagrangian relevant to the interaction of few-nucleon systems is
\begin{equation}
\mathcal{L}=\mathcal{L}_{1}+\mathcal{L}_{2}+\mathcal{L}_{3},
\label{eq:FVEFTLag}
\end{equation}
where the strong interactions between nucleons arise from 
\begin{align}
\mathcal{L}_{1}&=N^{\dagger}\left(i D_{0}+\frac{\mathbf{D}^{2}}{2 M_{N}}\right) N +\ldots,\\
\mathcal{L}_{2}&=-{C_S} \left(N^{T} P_{i} N\right)^{\dagger}\left(N^{T} P_{i} N\right)
-{C_T}\left(N^{T} \overline{P}_{a} N\right)^{\dagger}\left(N^{T} \overline{P}_{a} N\right) +\ldots,\\
    \mathcal{L}_{3}& = -\frac{D_0}{6}(N^\dagger N)^3 +\ldots,
\end{align}
where $P_i$ and $\overline{P}_a$ are the projectors defined in Eq.~\eqref{eq:projectors}, $M_N$ is the nucleon mass, and $C_S$, $C_T$, and $D_0$ are the relevant two and three-nucleon low-energy constants (LECs).
The two-nucleon couplings can also be expressed in terms of alternative LECs $C_{0,1}$ through the relations
\begin{equation}
C_T=C_{0}-3 C_{1} \text { and } C_S=C_{0}+C_{1}.
\end{equation}

The $n$-particle non-relativistic Hamiltonian corresponding to Eq.~\eqref{eq:FVEFTLag} is
\begin{equation}
H=-\frac{1}{2 M_N} \sum_{i} \nabla_{i}^{2}+\sum_{i<j} V_{2}\left({\bf r}_{i j}\right)+\sum_{i<j<k} V_{3}\left({\bf r}_{i j}, {\bf r}_{j k}\right),
\label{eq:Hamiltonian}
\end{equation}
where the integers $\{i,j,k\}$ label the particle, ${\bf r}_{ij} ={\bf r}_i - {\bf r}_j$ denotes the displacement between particles $i$ and $j$, and the two and three-particle potentials are given by
\begin{equation}
V_{2}\left({\bf r}_{i j}\right)=\left(C_{0}+C_{1} \sigma_{i} \cdot \sigma_{j}\right) g_{\Lambda}\left({\bf r}_{i j},L\right), 
\hspace{5mm}
V_{3}\left({\bf r}_{i j}, {\bf r}_{j k}\right)=D_0 \sum_{c y c} g_{\Lambda}\left({\bf r}_{i j},L\right) g_{\Lambda}\left({\bf r}_{j k},L\right).
\end{equation}
Here $g_\Lambda({\bf r},L)$ includes the Gaussian smearing which is used to regulate the interactions, and periodicity in the finite spatial volume of extent $L$ has been imposed in the regulator:
\begin{equation}
    g_\Lambda({\bf r},L) =\frac{\Lambda^{3}}{8 \pi^{3 / 2}} \prod_{\alpha=x,y,z} \sum_{q^{(\alpha)}=-\infty}^\infty \exp \left(-\Lambda^{2} (r_\alpha-L q^{(\alpha)})^{2} / 4\right),
\label{eq:periodic_regulator}
\end{equation}
where ${\bf r}=(r_x,r_y,r_z)$.

As described in Refs.~\cite{Eliyahu:2019nkz,WP}, the stochastic variational approach proceeds by the optimization of a two or three-body variational wavefunction defined in terms of correlated Gaussian basis components to minimize the expectation value of the Hamiltonian in Eq.~\eqref{eq:Hamiltonian} and converge to a representation of the ground-state wavefunction. Since rotational symmetry is broken by the lattice geometry, shifted correlated Gaussians are used  \cite{Eliyahu:2019nkz}. Defining a trial wavefunction as a linear combination of such shifted correlated Gaussians, the linear coefficients of the terms are optimized by solving  the  generalized  eigenvalue  problem  of the variational method; the approach taken to optimization is as detailed in Ref.~\cite{WP}. Given wavefunctions optimized in the same  volumes as the LQCD calculations, the LECs $C_S$, $C_T$, and $D_0$ of the pionless EFT Lagrangian can be constrained by matching the finite volume energies to the LQCD results, with the allowed range of LECs determined by a fit to the constraints from the three volumes. With the LECs fixed, volume-extrapolated energies are obtained using wavefunctions optimized at infinite volume.

\begin{figure}[!t]
	\includegraphics[width=0.475\columnwidth]{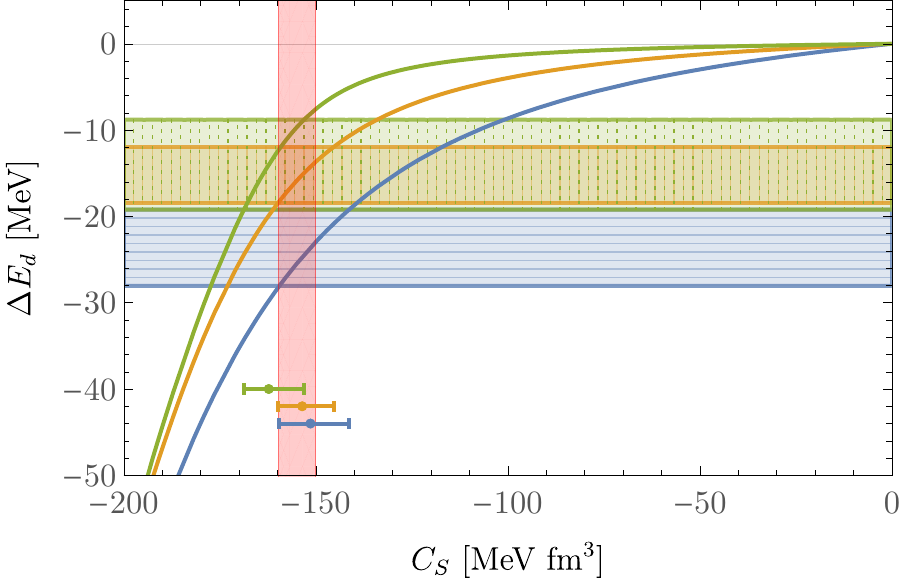}
	\includegraphics[width=0.475\columnwidth]{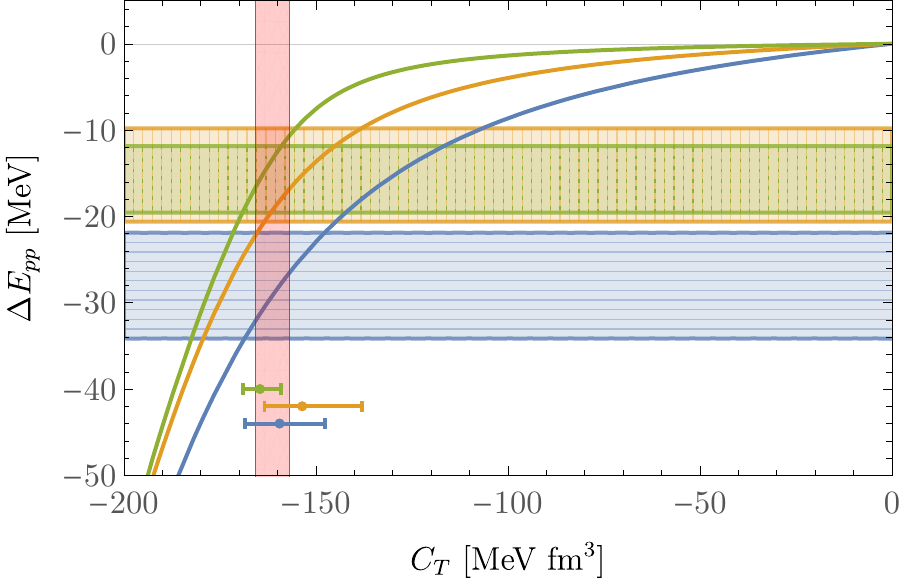}
	\caption{\label{fig:2bodyfits} Determination of the EFT two-body couplings, $C_{S,T}$ from the deuteron (left) and $pp$ (right) energies. The horizontal bands correspond to the energy shifts from the LQCD calculations on the E24 (blue), E32 (orange), and E48 (green) ensembles. The curves in each figure show the FVEFT two-body energy shifts for $L\in\{24,32,48\}$ (using the same color scheme) for the EFT cutoff parameter $r_0=0.2$ fm~\cite{WP}. 
	The LECs are determined by a simultaneous optimization matching the LQCD constraints from all three volumes to the corresponding curves; the results are shown as the red bands. To guide the eye, the intercept ranges of each band with the corresponding curve are shown as the colored bars.
	}
\end{figure}

\begin{figure}[!t]
	\includegraphics[width=0.6\columnwidth]{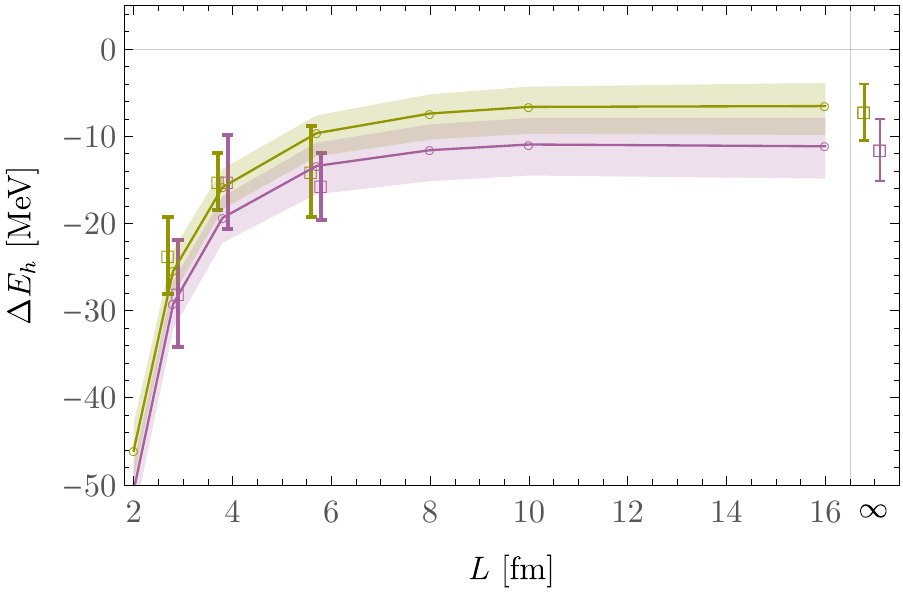}
	\caption{\label{fig:2bodyextraps} Extrapolation of the deuteron (green) and $pp$ (purple) energies to the infinite volume limit. Data points denote lattice results (from Ref.~\cite{Orginos:2015aya}), displaced slightly horizontally for clarity, while the shaded bands and infinite-volume points show the result of the FVEFT extrapolation.}
\end{figure}

\begin{figure}[!th]
	\includegraphics[width=0.6\columnwidth]{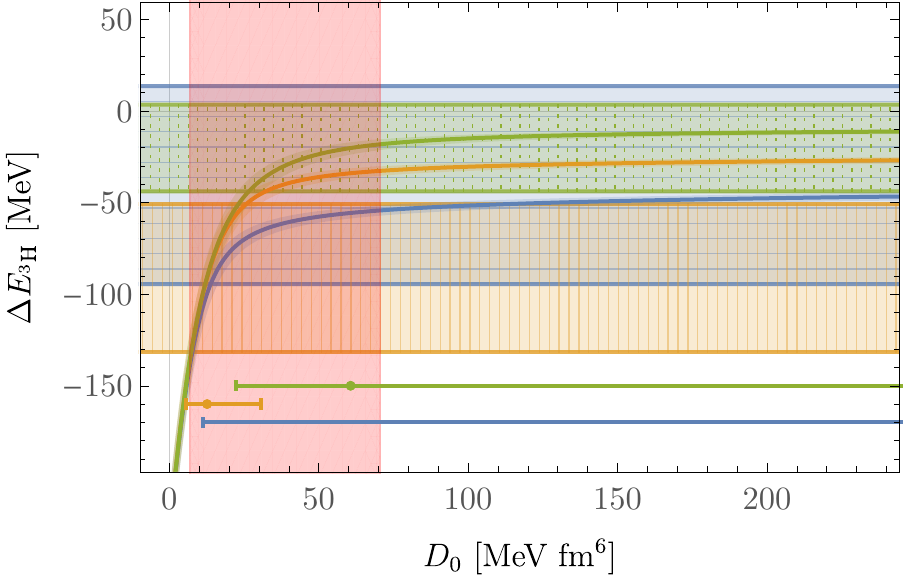}
	\caption{\label{fig:BtvsD0} Determination of the EFT three-body couplings $D_0$. As in Fig.~\ref{fig:2bodyfits}, the horizontal bands correspond to the energy shifts from the LQCD calculations on the E24 (blue), E32 (orange), and E48 (green) ensembles. The curves in each figure show the corresponding FVEFT energy shifts for the EFT cutoff parameter $r_0=0.2$ fm. The uncertainties on the curves are propagated from the uncertainties on the two-body LECs.}
\end{figure}

Figure~\ref{fig:2bodyfits} shows the determination of the two-body LECs from the LQCD results for the deuteron and dineutron energy shifts in the three lattice volumes.  The couplings are regulator-scale--dependent but the resulting energy shifts are not; calculations with cutoff parameter $\Lambda=\sqrt{2}/r_0$ with $r_0\in\{0.3,0.4\}$ fm result in indistinguishable results from those with $r_0=0.2$ fm which are shown here. 
The extrapolation to infinite volume using these couplings is shown in Fig.~\ref{fig:2bodyextraps}. 
Figure \ref{fig:BtvsD0} shows the same analysis of the three body system, determining the three-nucleon coupling $D_0$ and leading to the infinite-volume extrapolation presented in Fig.~\ref{fig:MpvsL}.%

Using the optimized wavefunctions, matrix elements of the isovector axial-vector current can be computed for the same finite volume as the LQCD calculations to fix the corresponding LECs, and the extrapolation to infinite-volume can then be undertaken in the same manner as that used for the binding energies themselves.
The isovector axial-vector current is expressed in pionless EFT as given in Eq.~\eqref{eq:axialcurrent} in the main text. In position space, the two-nucleon contribution proportional to $L_{1,A}$ is implemented using the same gaussian regulator as for the potential, specifically
\begin{align}
L_{1, A}\left[\left(N^{T}({\bf r}_i) P_{i} N({\bf r}_i)\right)^{\dagger}\left(N({\bf r}_j)^{T} \bar{P}_{a} N({\bf r}_j)\right)+\text{h.c.}\right] g_\Lambda({\bf r}_{ij})\,.
\end{align}
The optimised triton wavefunction corresponding to volume of  the E32 ensemble is used to compute the finite-volume axial matrix element in Eq.~\eqref{eq:EFTdude}. Approximating the 
 triton wavefunction as a tensor product
\begin{align}
    \left|\Psi_{\hthree},s\right\rangle = \left|\chi_{s}\right\rangle \otimes \left|\psi({\bf r}_1,{\bf r}_2,{\bf r}_3)\right\rangle,
\end{align}
the
computation of the matrix elements separates into the spin-isospin and spatial parts. The simplest spin-isospin wavefunction for the spin-up component is given by
\begin{align}
    \left|\chi_{s=+}\right\rangle =& \frac{1}{\sqrt{6}}\left[
    \ket{n^\uparrow p^\uparrow n^\downarrow} - \ket{ n^\downarrow p^\uparrow n^\uparrow}
     -\ket{p^\uparrow n^\uparrow n^\downarrow}\right. \left. + \ket{ p^\uparrow n^\downarrow n^\uparrow}
    - \ket{n^\uparrow n^\downarrow p^\uparrow } + \ket{ n^\downarrow n^\uparrow p^\uparrow}
    \right],
\end{align}
with an analogous expression for the spin-down wavefunction.
The spatial part of the matrix element is determined from the variationally-optimized triton  wavefunction and is given by
\begin{equation}
      h_{\hthree}(\Lambda,L)=\frac{\int \prod_kd^3{\bf r}_k \sum_{i< j} g_\Lambda({\bf r}_{ij}) |\psi({\bf r}_1,{\bf r}_2, {\bf r}_3)|^2}{\int \prod_kd^3{\bf r}_k|\psi({\bf r}_1,{\bf r}_2, {\bf r}_3)|^2} \,,
\end{equation}
where $\{i,j,k\} \in \{1,2,3\}$ and $\psi({\bf r}_1,{\bf r}_2, {\bf r}_3)$ is the position-space representation of the  spatial wavefunction of the triton.

The LQCD ratio of the triton to proton axial couplings is reproduced by tuning the LEC ratio $\widetilde{L}_{1,A}=L_{1,A}/g_A$ in Eq.~\eqref{eq:EFTdude} as shown in Fig.~\ref{fig:effchargeatiovsL}.
Having fixed this LEC ratio, the infinite-volume matrix element is evaluated using the infinite-volume variational wavefunction, and the volume-dependence is evaluated using wavefunctions optimized at a range of intermediate volumes. Note that just as for the LECs $\{C_S,C_T,D_0\}$, the LEC ratio $\widetilde{L}_{1,A}$ is determined in the exponential regulator scheme, but the evaluated axial matrix element itself is independent of the choice of regulator.

\bibliography{tritongA}

\end{document}